\documentclass[letterpaper,journal]{IEEEtran}
\IEEEoverridecommandlockouts

\AtBeginDocument{%
  \providecommand\BibTeX{{%
    \normalfont B\kern-0.5em{\scshape i\kern-0.25em b}\kern-0.8em\TeX}}}

\usepackage{amsmath,amsfonts}
 
\usepackage{amssymb}

\usepackage[compatibility=false]{caption}
\usepackage{subcaption}
\usepackage{stfloats}
\usepackage{marvosym}
\usepackage{balance}
\usepackage{graphicx}
\usepackage{tikz}
\usetikzlibrary{shapes.geometric}
\usepackage{pgfplots}
\pgfplotsset{compat=1.18}
\usepackage{seqsplit}
\usepackage{array}
\usepackage{booktabs} 
\usepackage[normalem]{ulem}
\usepackage{xurl}
\usepackage{hyperref}
\usepackage{comment}
\usepackage{xcolor}
\usepackage{pifont} 
\usepackage[ruled,vlined,linesnumbered]{algorithm2e}

\usepackage{multirow}
\usepackage{makecell}
\usepackage{colortbl} 
\usepackage{adjustbox}

\definecolor{pmRed}{RGB}{218,42,42}
\definecolor{pmGreen}{RGB}{43,139,87}
\definecolor{pmGold}{RGB}{224,163,20}

\newcommand{\cmark}{\textcolor{pmGreen}{\ding{51}}}
\newcommand{\xmark}{\textcolor{pmRed}{\ding{55}}}
\newcommand{\underhead}[1]{\texorpdfstring{\uline{#1}}{#1}}
\newcommand{\pmfn}[1]{\textcolor{violet}{\texttt{#1}}}

\newenvironment{packeditemize}{
	\begin{list}{$\bullet$}{
			\setlength{\labelwidth}{4pt}
			\setlength{\itemsep}{0pt}
			\setlength{\leftmargin}{\labelwidth}
			\addtolength{\leftmargin}{\labelsep}
			\setlength{\parindent}{0pt}
			\setlength{\listparindent}{\parindent}
			\setlength{\parsep}{0pt}
			\setlength{\topsep}{1pt}}}{\end{list}}

\begin{document}

\title{Deanonymizing Monero Transactions \\in Tor Network\thanks{
An abstract version was published in WWW'24 workshop~\cite{shi2024deanonymizing}.}}

\author{ 
 Ruisheng Shi$^1$,
 Shihan Zhang$^1$,
 Yulian Ge$^1$,
 Lina Lan$^1$,
 Qingfeng Zhang$^1$,
 Qin Wang$^2$ \\
\vspace{5pt}
\textit{$^1$Beijing University of Posts and Telecommunications} $|$ \textit{$^2$Independent}
}

\maketitle

\begin{abstract}
Monero is a privacy-focused cryptocurrency that deploys the Dandelion++ protocol and incorporates anonymity networks (such as Tor and I2P) to prevent malicious attackers from linking transactions with their source IPs. 
In this paper, we demonstrate that Monero's integration of the Tor network introduces a fundamental vulnerability: a Monero Tor node's originated transactions are exclusively forwarded to two outgoing Tor hidden service nodes (proxy nodes) prior to clearnet propagation, enabling an adversary to capture originated transactions by occupying the target node's outgoing connections. Based on this observation, we propose \textit{ProxyMark}, a three-stage deanonymization framework for the Monero Tor network, comprising node role identification, originated transaction identification, and node location deanonymization. Through experiments on the live Tor network, Monero mainnet, and testnet, we empirically demonstrate the effectiveness of \textit{ProxyMark} in successfully deanonymizing transactions originating from Monero nodes over Tor.
\end{abstract}


\section{Introduction}\label{introduction}

Monero is considered one of the most representative anonymous cryptocurrencies due to the application of multiple privacy-preserving technologies. As of April 15th, 2026, Monero's market capitalization is \$6.32 billion \cite{CoinMarketCap}, ranking it first among privacy-focused cryptocurrencies. The privacy features of Monero have led to its widespread use in darknet markets, earning it the title of the king of darknet.

In Monero peer-to-peer (P2P) network, transactions are forwarded by nodes to their neighbors after created, enabling all nodes in the P2P network to receive transactions. De-anonymisation attacks on the network layer aim to obtain source IPs of transactions by analyzing the propagation path of transactions in the P2P network. Existing research on network-layer transaction de-anonymization for cryptocurrencies \cite{koshy2014,biryukov2014,Biryukov2015,tramer2020,apostolaki2021,gao2021,zhang2026deanonymizing} is mostly proposed for Bitcoin's diffusion transaction propagation mechanism. However, since Monero uses a transaction propagation mechanism different from Bitcoin, existing research methods cannot successfully de-anonymize Monero transactions. It is necessary to propose a de-anonymization approach aiming at Monero transaction propagation mechanism.

Monero conceals the source IP address of transactions through Dandelion++ \cite{fanti2018dandelion++} and anonymity networks~\cite{MoneroAnonymityNetworks}. Dandelion++ divides transaction propagation into two phases: stem and fluff. In the stem phase, a node forwards a transaction to one randomly selected outgoing neighbor; after a random number of hops, the transaction enters the fluff phase and is broadcast to all neighbors with random delays. An observer that receives the transaction cannot directly infer how many stem hops have already occurred.

\smallskip
\noindent\textbf{Monero over Tor.} Monero also supports routing P2P traffic through Tor and I2P. When Monero uses Tor, P2P messages are relayed through encrypted Tor circuits, hiding the node's IP address from its Monero peers and hiding the communication relationship from local network observers. In practice, Monero over Tor creates two main types of participants. A Tor client node uses Tor to establish outbound connections and may connect either to clearnet public peers through Tor exits or to Monero hidden service peers inside Tor. A Monero Tor hidden service node publishes an onion address, accepts incoming Tor connections, and can also maintain outbound connections to public peers and hidden service peers.

This distinction matters because Tor is not only a transport wrapper around the ordinary Monero P2P protocol. Monero treats public peers and hidden service peers differently for block synchronization, address propagation, and transaction forwarding. In particular, hidden service connections are used as a protected forwarding layer for transactions originated by Monero nodes over Tor, while public peers remain necessary for clearnet propagation and blockchain synchronization. As a result, Tor hides node locations but also introduces role dependent identifiers, peer lists, and forwarding paths that are absent from ordinary clearnet Monero nodes. This mismatch between Monero's transaction logic and Tor's connection model is the entry point of our analysis.

\smallskip
\noindent\textbf{Intuition for deanonymizing}. Through a systematic analysis of Monero Tor nodes' transaction propagation behavior, we find that, by default, a Monero Tor node's originated transactions are first forwarded to two outgoing Tor hidden service nodes, which we refer to hereafter as proxy nodes. These proxy nodes form a small protocol-defined choke point: before an originated transaction enters clearnet diffusion, it must pass through one of the target's selected hidden-service peers. The proxy nodes then relay the transactions to clearnet nodes via the Dandelion++ protocol, after which propagation proceeds entirely within the clearnet under Dandelion++. In contrast, relayed transactions are forwarded directly to clearnet neighbors. As a result, the direction and neighbor type of a received transaction carry useful information. If an adversarial hidden service peer becomes one of the target's outgoing proxy nodes, a transaction arriving from the target over that hidden-service connection is not merely another relayed transaction; under Monero's forwarding rules, it is generated by the peer on the other side of the connection. The attack intuition is therefore to bias peer selection so adversarial hidden-service nodes become proxy nodes, use those connections to capture originated transactions, and then bind the observed Tor-level peer identifier to a real IP address through traffic watermarking. This does not require breaking Tor encryption or Monero cryptography; it exploits the boundary where Monero's forwarding policy assigns different roles to public and hidden-service connections.

\smallskip
\noindent\textbf{Technical challenges.} Thus, deanonymising transactions in Monero Tor networks face three technical challenges (TCs).

\begin{packeditemize}
\item \textit{TC-I: Occupying outgoing connections.} Capturing originated transactions requires adversarial hidden service nodes to be selected among the target's outgoing hidden-service peers. This is nontrivial because Monero Tor nodes maintain a bounded set of outgoing hidden-service connections, choose replacements from locally maintained graylists and whitelists, and expose different reachability properties for hidden-service nodes and client nodes. The adversary must therefore influence the target's future peer choices without relying on direct inbound access to Tor clients or on disruptive eclipse-style assumptions. Existing Sybil or DoS-based approaches either require continuously increasing resources or depend on fragile Tor vulnerabilities~\cite{biryukov2013trawling,tan2018toward,zhang2024hsdirsniper}.

\item \textit{TC-II: Biasing proxy selection.} Even after occupying outgoing connections, the adversary captures a transaction only when one occupied peer is selected as one of the proxy nodes for the target's originated traffic. Monero does not choose proxies uniformly from all neighbors: it filters candidates using blockchain-height information and periodically refreshes the selected proxies. Thus, connection occupation alone is insufficient; adversarial peers must remain eligible and appear fresher than benign peers while still behaving like plausible Monero hidden-service neighbors.

\item \textit{TC-III: Linking identifiers to real IP addresses.} Captured transactions are still tied only to Tor-level identifiers, not real IP addresses. The identifier available to the adversary is role dependent: a hidden-service target naturally exposes an onion address, whereas a Tor client requires an assigned connection-specific identifier. Any IP-binding method must also tolerate encrypted Tor cells, Monero white-noise traffic, protocol-required Timed Sync messages, and Tor flow-control cells. Existing address-cookie and watermarking ideas~\cite{Biryukov2015,gao2021} therefore do not directly fit Monero because of unconditional address sharing, different P2P messages, traffic noise, and distinct hidden-service/client roles.

\end{packeditemize}

\smallskip
\noindent\textbf{Our solution: \textit{ProxyMark}.} To address these challenges, we propose \textit{ProxyMark}, a transaction deanonymization framework for the Monero Tor network. \textit{ProxyMark} chains together transaction capture and node-location deanonymization: it first learns the role and identifier of a Tor-side Monero peer, then increases the chance that originated transactions are delivered to adversarial hidden service peers, and finally associates the resulting identifier with a real IP address.

The framework has three stages. First, \textit{node role identification} exploits a difference in Timed Sync Response peer lists: hidden service nodes repeatedly advertise their own onion address, whereas Tor client nodes only forward learned addresses. This step determines whether the target is a hidden service or client node and, for hidden services, extracts its onion address. Second, \textit{originated transaction identification} actively fills target peer lists with adversarial onion addresses to occupy outgoing hidden-service connections (TC-I), then advertises falsified fresh block heights so that occupied peers are more likely to be selected as proxy nodes (TC-II). Third, \textit{node location deanonymization} assigns role-specific identifiers and embeds them into traffic as signal watermarks. A malicious Tor relay that observes the marked circuit can recover the identifier and bind it to the source IP address, while role-specific embedding and denoising handle Monero white noise and Tor control traffic (TC-III).

\smallskip
\noindent\textbf{Contributions.} The main contributions of our work are: 
\begin{packeditemize}

\item We present \textit{ProxyMark}, a three-stage framework for deanonymizing Monero transactions sent over Tor. Rather than attacking Tor encryption, \textit{ProxyMark} exploits Monero-over-Tor P2P behavior to identify Tor-level node identifiers, capture originated transactions, and link the identifiers to source IP addresses through traffic watermarking. \textit{ProxyMark} includes three new designs:

\begin{packeditemize}
    \item An \textit{onion address analysis} method to identify Monero Tor hidden service nodes and extract their onion addresses by exploiting deterministic differences in address propagation between hidden service nodes and Tor client nodes.

    \item An \textit{originated transaction capture} method combining \textit{connection occupation} and \textit{proxy selection bias} to increase the adversary's share of outgoing hidden-service peers and increse the likelihood of being selected as proxy nodes.

    \item A \textit{role-specific watermarking} method to link Monero Tor identifiers to source IP addresses. The method embeds identifiers through Monero P2P message timing and uses role-specific denoising to handle Monero white-noise traffic, protocol messages, and Tor flow-control cells.

\end{packeditemize}

    \item We validate \textit{ProxyMark} on the live Tor network, Monero mainnet, and testnet. Onion Address Analysis achieves 100\% precision and recall. Connection Occupation captures 7--11 of 12 outbound hidden-service connections after restarts and 8--10 of 10 after periodic replacement. With one occupied connection, Proxy Selection Bias raises the proxy-selection rate from 15.3\% to 35.7\% (+20.4 percentage points). Role-Specific Watermarking achieves 100\% precision with 93.8\% recall for hidden service nodes and 100\% precision with 91.4\% recall for Tor client nodes.

    \item We analyze the root causes of the identified leakage and discuss protocol-level mitigations for each attack stage, including changes to onion-address advertisement, transaction forwarding over anonymity networks, peer selection, and message-rate handling.

\end{packeditemize}

\section{Dissecting Monero Tor Network}
\label{sec-bck}

This section summarizes the Monero and Tor behaviors used by \textit{ProxyMark}. We distinguish ordinary nodes, Tor client nodes, and Tor hidden service nodes, and use \emph{Monero nodes over internal Tor} for Tor client or hidden service nodes communicating with Monero hidden service peers\footnote{Traffic between a Tor client and Tor hidden services remains inside Tor, whereas traffic from a Tor client to public services exits Tor before reaching.}.

\newsavebox{\tmpbox}
\newlength{\firstrowheight}
\begin{figure*}[t]
    \centering
    \savebox{\tmpbox}{\includegraphics[width=0.25\textwidth]{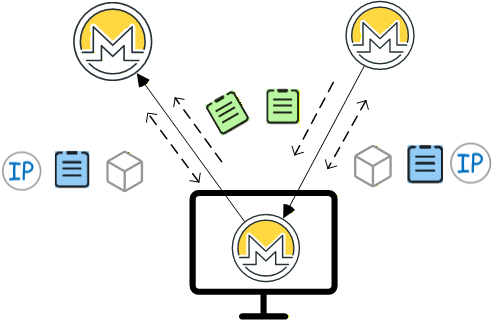}}
    \settoheight{\firstrowheight}{\usebox{\tmpbox}}
    
    \begin{subfigure}[t]{0.25\textwidth}
        \usebox{\tmpbox}
        \caption{}
        \label{img_nodes_connection_ordinary}
    \end{subfigure}\hfil
    \begin{subfigure}[t]{0.25\textwidth}
        \includegraphics[width=\linewidth]{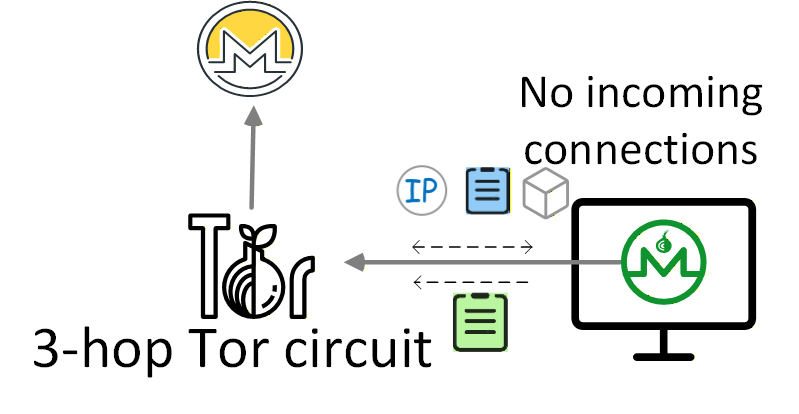}
        \caption{}
        \label{img_nodes_connection_tor_client_to_only_public}
    \end{subfigure}\hfil
    \begin{subfigure}[t]{0.25\textwidth}
        \includegraphics[width=\linewidth]{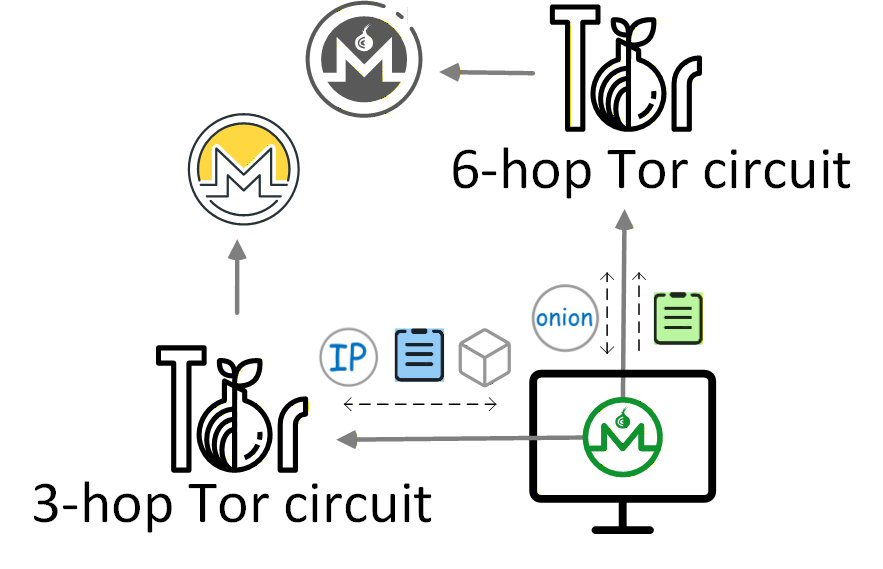}
        \caption{}
        \label{img_nodes_connection_tor_client_to_Tor_to_public_and_hs}
    \end{subfigure}\hfil
    \begin{subfigure}[t]{0.2\textwidth}
        \includegraphics[height=\firstrowheight, keepaspectratio]{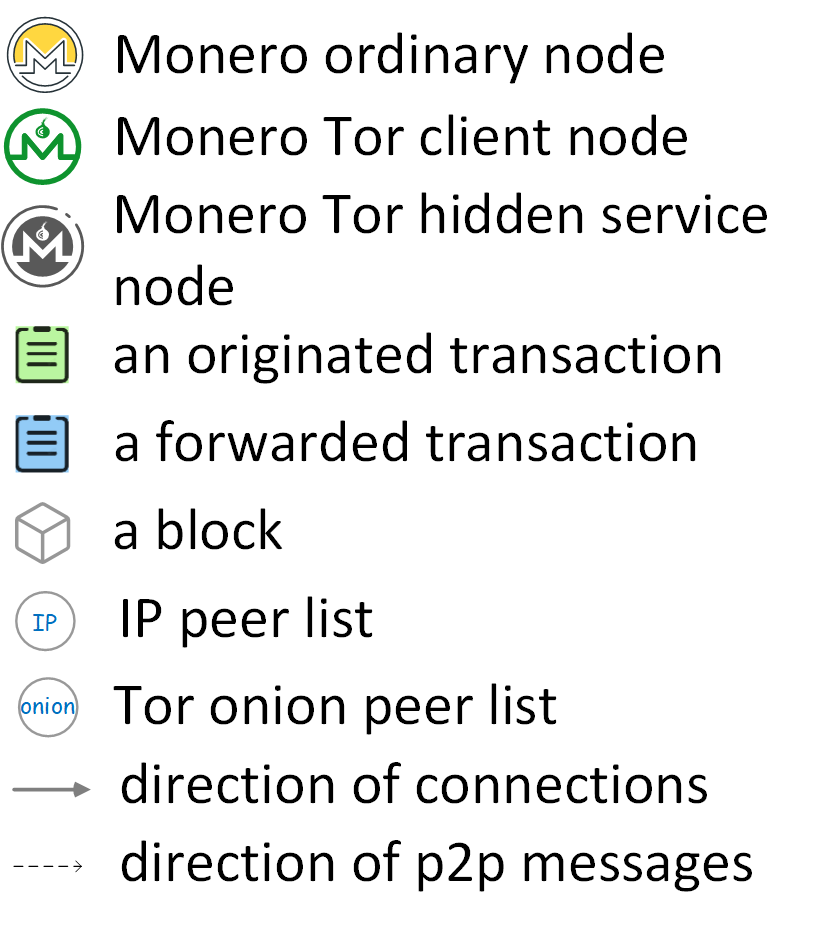}
    \end{subfigure}

    \medskip
    \begin{subfigure}[t]{0.2\textwidth}
        \includegraphics[width=\linewidth]{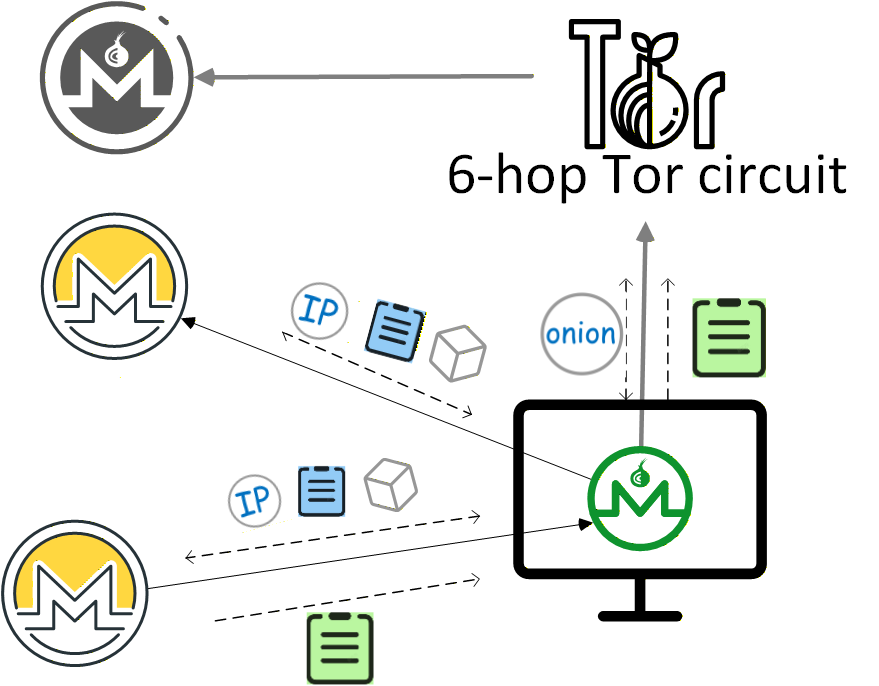}
        \caption{}
        \label{img_nodes_connection_tor_client_to_direct_public_and_hs}
    \end{subfigure}\hfil
    \begin{subfigure}[t]{0.30\textwidth}
    \includegraphics[width=\linewidth]{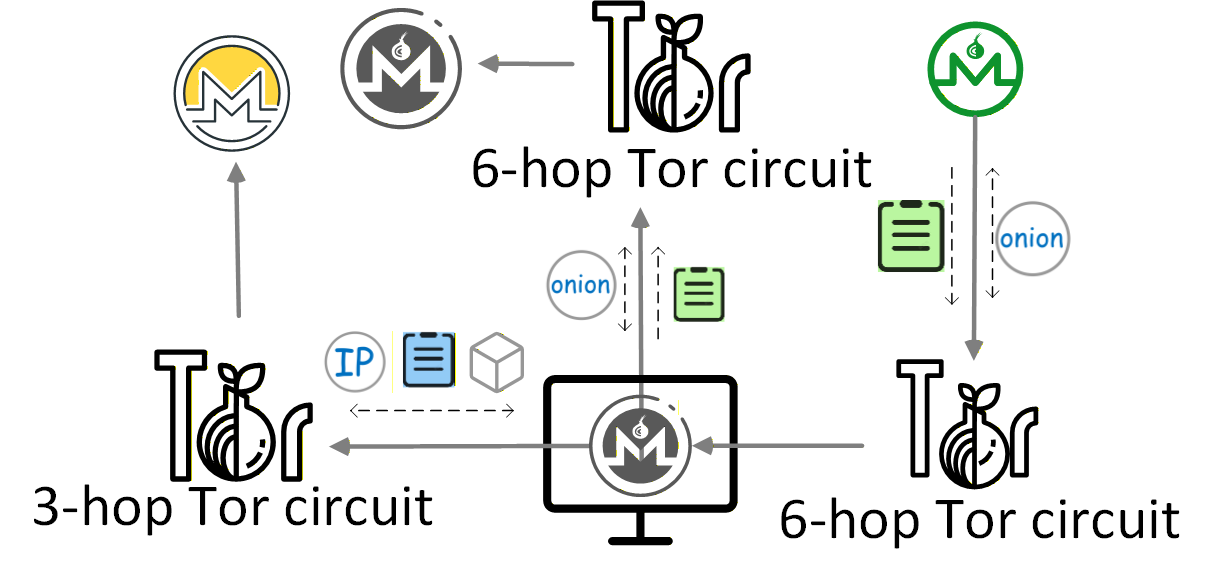}
        \caption{}
        \label{img_nodes_connection_hs_to_Tor_to_public_and_hs}
    \end{subfigure}\hfil
    \begin{subfigure}[t]{0.30\textwidth}
        \includegraphics[width=\linewidth]{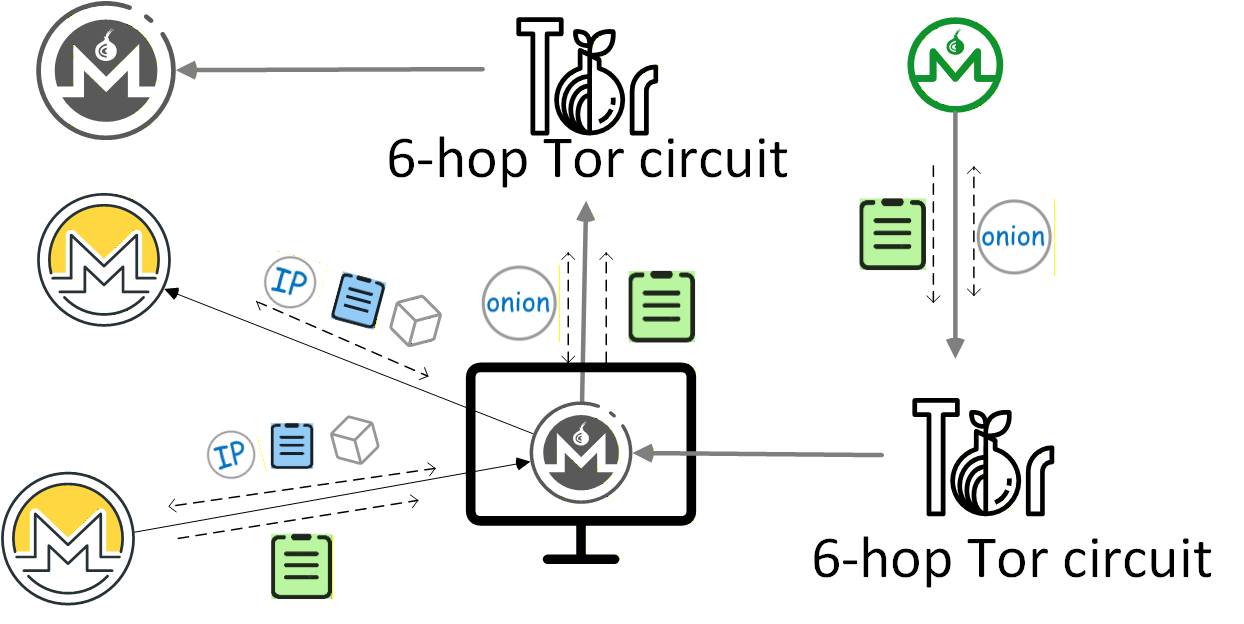}
        \caption{}
        \label{img_nodes_connection_hs_to_direct_public_and_hs}
    \end{subfigure}
    
    \caption{Connection patterns of Monero nodes: (a) ordinary node; (b) Tor client connected only to public peers; (c) Tor client connected to public peers through Tor and hidden service peers; (d) Tor client connected directly to public peers and to hidden service peers; (e) hidden service node connected to public peers through Tor and hidden service peers; (f) hidden service node connected directly to public peers and to hidden service peers. Tor circuit structures are shown in Appendix~\ref{appendix_tor_circuit_construction}.} 
    \label{img_nodes_connection}
\end{figure*}

\subsection{Monero's P2P Protocol Messages}
\label{subsec:p2p_messages}
Three types of P2P messages are relevant to the node behaviors discussed below: Handshake messages (for connection establishment, exchanging node configuration, blockchain synchronization status, and a peer list), Timed Sync messages (for periodic neighbor updates on online status, blockchain height, and newly connected peers), and New Transactions messages (for transaction propagation). On hidden-service peer links, block-related messages are excluded to prevent sybil attacks. In normal protocol operation, Timed Sync and New Transactions messages are exchanged after the Handshake exchange is complete. The initialization process of Monero nodes is provided in Appendix~\ref{appendix_initialization_process_of_Monero_nodes}.

\subsection{Tor Network and Tor Hidden Service}
\label{subsec:tor_network_hs}
The second-generation onion router (Tor) is a widely used anonymity network with over 7000 volunteer-operated relays. When a Tor client communicates, the onion proxy (OP) selects a circuit of typically three relays (entry, middle, exit), negotiates a session key with each, and encrypts messages layer-by-layer for decryption at each hop, hiding both communication endpoints. Tor uses fixed-size (514-byte) cells to reduce size-based traffic analysis; Appendix~\ref{appendix_tor_cell} gives cell details.

To protect service providers, a Tor client and a Tor hidden service establish separate circuits to a rendezvous point, concealing both IP addresses. To mitigate anonymity attacks, each Tor user selects two guard-flagged relays, uses one as the circuit entry relay, and rotates the pair approximately every 120 days~\cite{TorForum,NumEntryGuards}.

\subsection{Monero Ordinary Nodes}

As shown in Figure~\ref{img_nodes_connection_ordinary}, a Monero ordinary node maintains 12 outgoing connections to public nodes and, by default, accepts unlimited incoming public connections. It propagates transactions according to Dandelion++~\cite{fanti2018dandelion++} and synchronizes blocks with both incoming and outgoing public peers. For address propagation, an outgoing public peer returns up to 250 known IP addresses during the handshake, and neighbors later exchange additional IP addresses through periodic Timed Sync messages.

This ordinary-node behavior serves as the baseline for the Tor-specific cases below. Monero Tor nodes still rely on public peers for block synchronization and final clearnet transaction propagation; the attack-relevant differences arise from the additional hidden-service peer lists, forwarding paths, and identifiers used on Tor-side connections.

\subsection{Monero Tor Client Nodes}
Since Monero nodes synchronize blocks only with public peers, a Monero node over internal Tor must connect to at least some public nodes. Accordingly, Monero Tor client nodes either connect only to public nodes, or connect to both public nodes and Monero Tor hidden service nodes.

\subsubsection[Public-Only Clients]{\underhead{Public-Only Clients}}
As shown in Figure~\ref{img_nodes_connection_tor_client_to_only_public}, this type has 12 outgoing connections to public nodes via Tor and no incoming connections. Its transaction propagation, address propagation, and blockchain synchronization behavior matches ordinary nodes, except that the public-peer connections are carried over Tor. Because it does not maintain Monero hidden-service peers, this type does not expose the hidden-service peer lists or proxy-node forwarding path used by \textit{ProxyMark}; we include it only to separate Tor transport from Monero's internal-Tor behavior.

\subsubsection[Clients with Hidden-Service Peers]{\underhead{Clients with Hidden-Service Peers}}
These nodes connect to public nodes either over Tor (Figure~\ref{img_nodes_connection_tor_client_to_Tor_to_public_and_hs}) or directly (Figure~\ref{img_nodes_connection_tor_client_to_direct_public_and_hs}). This case is also the outgoing-side baseline for Monero Tor hidden service nodes, because both maintain public peers for block synchronization and hidden-service peers for Tor-side transaction forwarding and onion-address exchange. Section~\ref{subsec:monero_tor_hs_nodes} therefore focuses only on what hidden service nodes add beyond this baseline.

\smallskip
\noindent\textbf{Connection management.} Such a node maintains 12 outgoing connections to public nodes and 12 to Monero Tor hidden service nodes. Nodes connecting directly to public nodes also accept an unlimited number of incoming public connections by default. Thus, hidden-service connections supplement public-peer connectivity: they carry onion-address exchange and transaction forwarding, while public peers remain necessary for block synchronization.

\smallskip
\noindent\textbf{Transaction propagation.} Sends forwarded transactions to public peers and sends originated transactions exclusively to outgoing Monero Tor hidden service peers. Nodes connecting directly to public nodes may receive transactions originated by their incoming public peers (Figure~\ref{img_nodes_connection_tor_client_to_direct_public_and_hs}).

\begin{figure}[pt]
  \centering
  \begin{subfigure}[pt]{0.4\textwidth}
      \includegraphics[width=\linewidth]{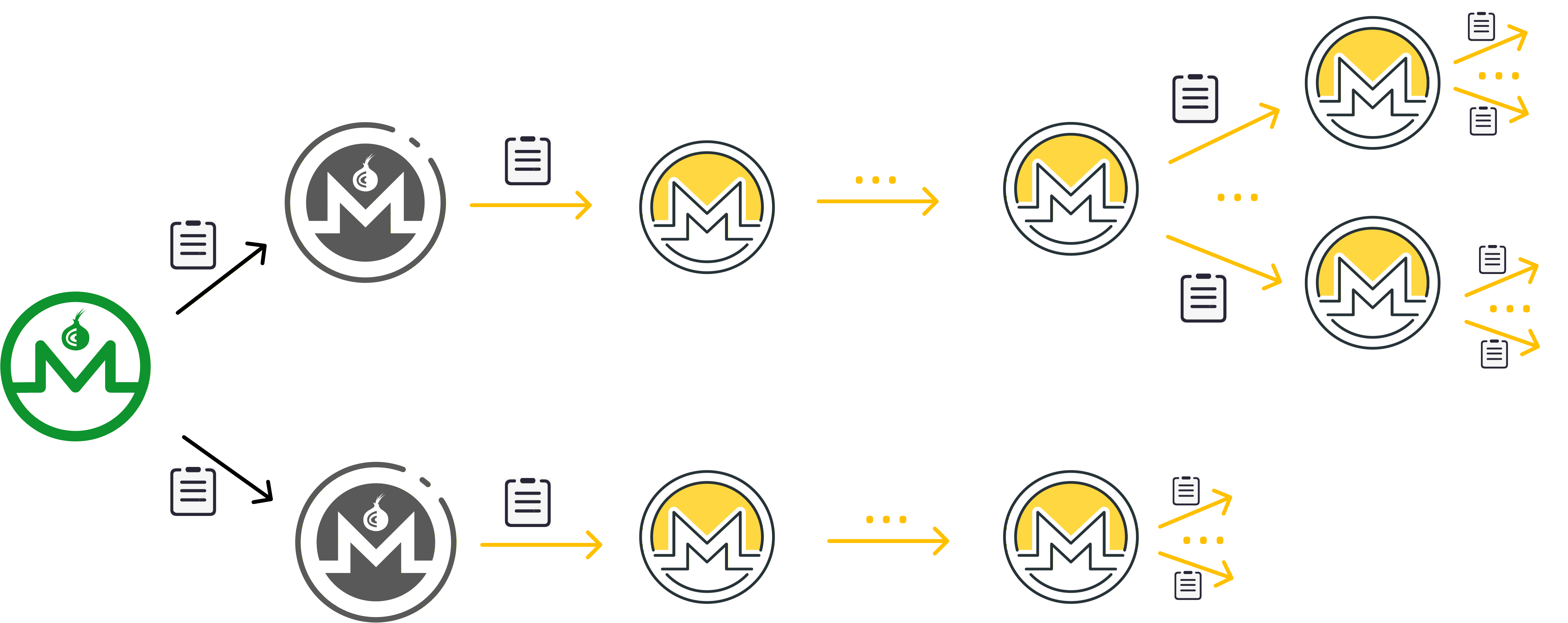}
      \caption{White noise enabled}
      \label{img_transaction-path_noise}
  \end{subfigure}
  \begin{subfigure}[pt]{0.4\textwidth}
        \includegraphics[width=\linewidth]{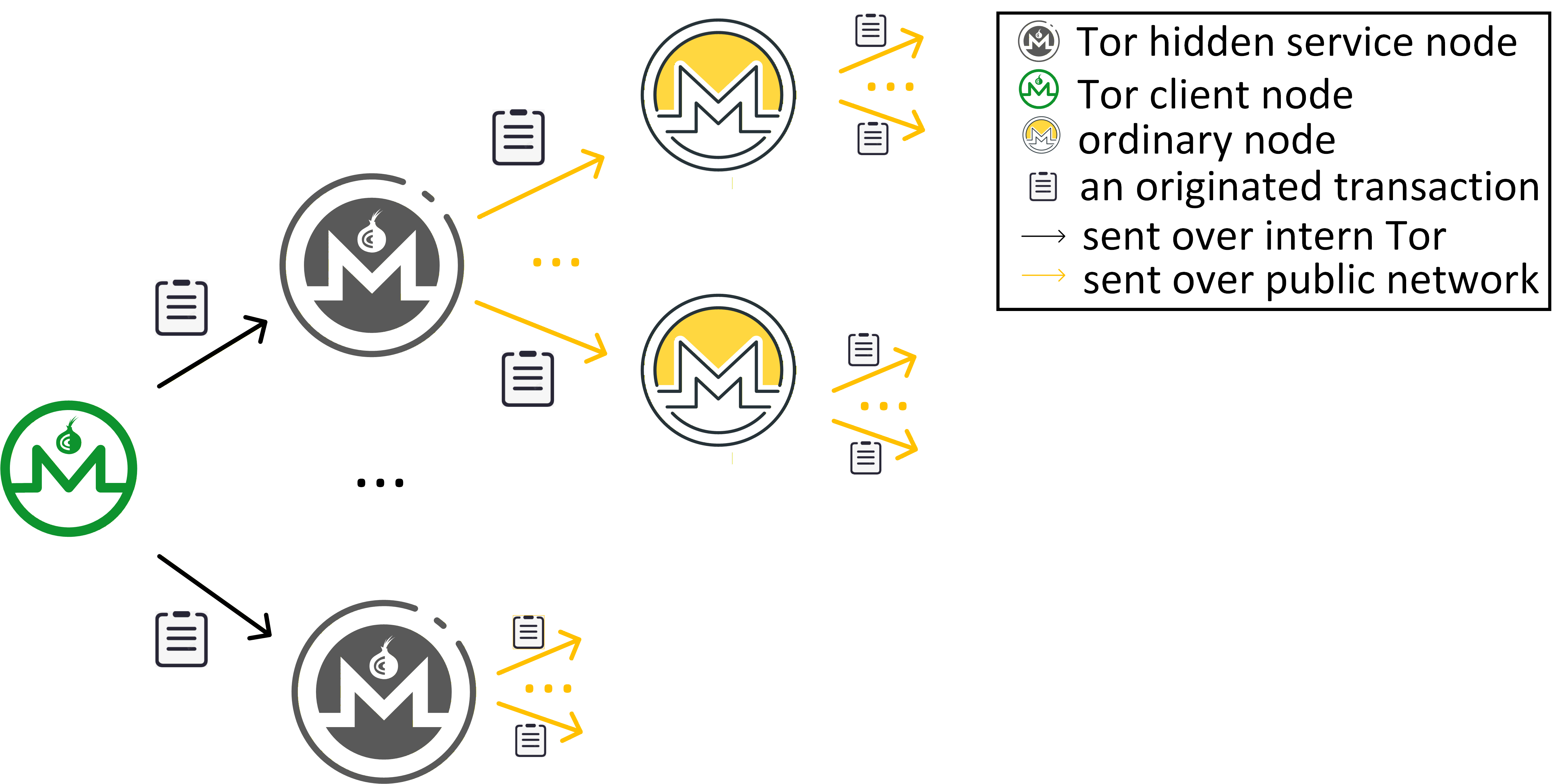}
      \caption{White noise disabled}
      \label{img_transaction-path_no_noise}
  \end{subfigure}
  \caption{Propagation path for transactions originated by Monero nodes over internal Tor.}
  \label{img_transaction-path}
\end{figure}

The propagation path is illustrated in Figure~\ref{img_transaction-path}. By default, originated transactions are sent first to two outgoing Monero Tor hidden service peers (proxy nodes), which then relay them to ordinary nodes via Dandelion++ for further clearnet propagation. These two proxy nodes are selected from outgoing Tor hidden service peers whose blockchain height meets a minimum threshold: if the node has completed synchronization, the threshold is its current blockchain height; otherwise, it is the greater of its current height and the median height of its outgoing Tor hidden service peers. This threshold thus generally reflects the latest blockchain height in the Monero network. The selected proxy nodes are re-randomized after a random interval (one epoch, from 5 to 5.5 minutes) or when a new outgoing connection to a Monero Tor hidden service node is established. These proxy-selection rules are the protocol basis for the connection-occupation and proxy-bias techniques in Section~\ref{subsec-originated-tx-identi}.

To prevent ISP attackers from correlating transactions with source IPs, two protection strategies are employed. First, to obscure transaction size and timing, the node sends fixed-size (3072-byte) white noise to its two outgoing Tor hidden service peers every 10 to 15 seconds; when an originated transaction is transmitted, it is padded to a multiple of this fixed size and sent in segments. Second, to prevent timing-based correlation, a random delay (Poisson-distributed with an average of 22.5 seconds) is introduced before the two proxy nodes forward transactions to ordinary nodes on the clearnet.

Monero Tor nodes may disable the above protection strategies, allowing their originated transactions to appear in the mempool immediately. In this case, the originated transaction is first broadcast to all outgoing Monero Tor hidden service peers with Dandelion++ fluff-phase delays, which then forward it to ordinary nodes for further propagation across the public network under the same fluff-phase rules.

\smallskip
\noindent\textbf{Address propagation.} Exchanges IP addresses with public peers and onion addresses with outgoing Monero Tor hidden service peers. Upon connection, the node receives 250 IP addresses from an outgoing public peer and 250 onion addresses from an outgoing Monero Tor hidden service peer. 

Each such node sends a Timed Sync Request to all neighbors every minute. In response, public peers return up to 250 IP addresses successfully connected to and not previously shared with the requester, and outgoing Tor hidden service peers return up to 250 onion addresses. Symmetrically, this node responds to Timed Sync Requests from public peers with up to 250 such IP addresses, and to those from outgoing Tor hidden service peers with up to 250 onion addresses.

\subsection{Monero Tor Hidden Service Nodes}
\label{subsec:monero_tor_hs_nodes}
Monero Tor hidden service nodes keep the same outgoing public-peer and hidden-service-peer behavior described above, but additionally publish an onion address and accept incoming connections from Monero nodes over internal Tor. They may connect to public peers via Tor (Figure~\ref{img_nodes_connection_hs_to_Tor_to_public_and_hs}) or directly (Figure~\ref{img_nodes_connection_hs_to_direct_public_and_hs}); nodes that connect directly to public peers also accept incoming public connections by default.

The incoming Tor side is the main role-specific difference. A hidden service node can receive transactions from incoming Monero Tor peers, while its own originated transactions still follow the proxy-node path in Figure~\ref{img_transaction-path}. This reachability lets an adversarial Monero peer probe a hidden service node and later inject role-specific watermarks.

\begin{figure}[!htbp]
  \centering
  \begin{subfigure}[pt]{0.4\textwidth}
      \includegraphics[width=\linewidth]{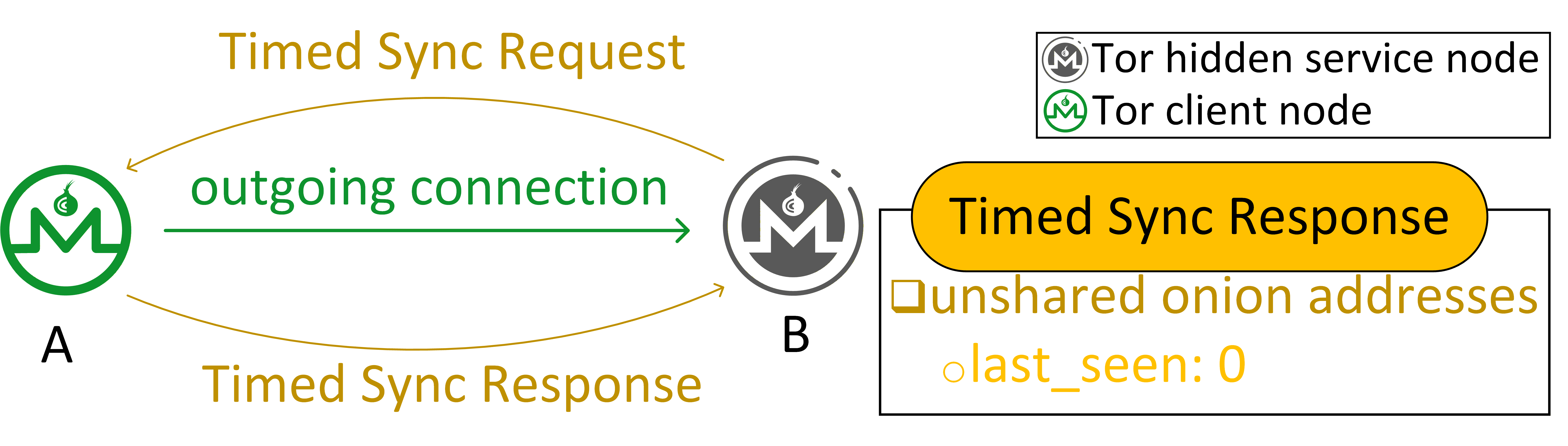}
      \caption{Tor client response}
      \label{img_onion_analysis_client}
  \end{subfigure}
  \begin{subfigure}[pt]{0.4\textwidth}
        \includegraphics[width=\linewidth]{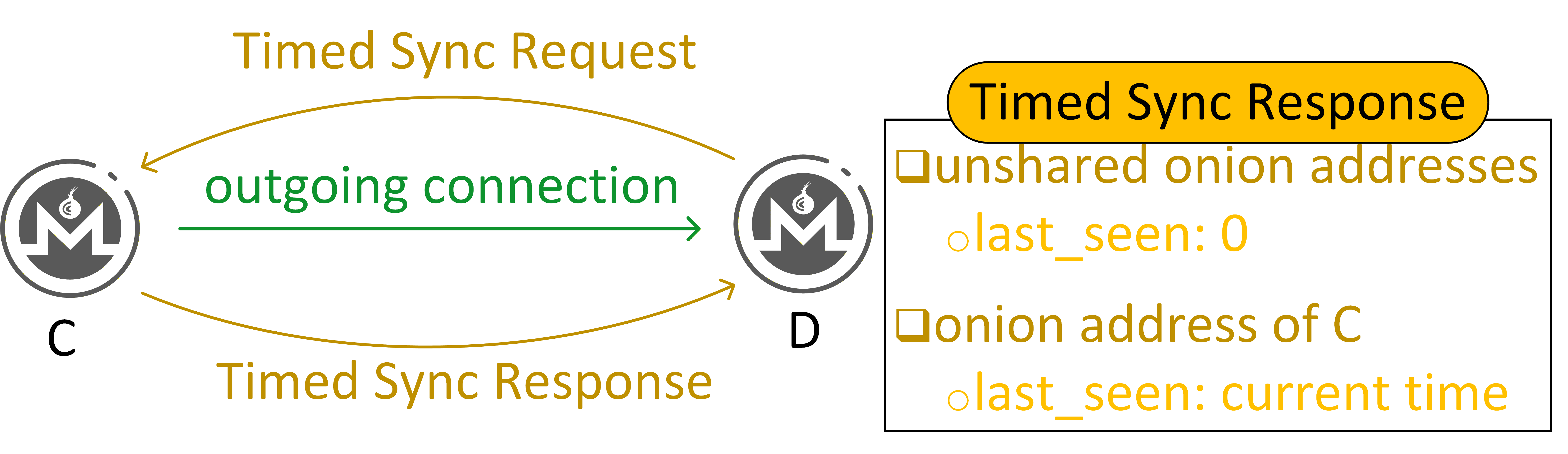}
      \caption{Hidden service response}
      \label{img_onion_analysis_hs}
  \end{subfigure}
  \caption{Onion-address lists in Timed Sync Response messages.}
  \label{img_onion_analysis}
\end{figure}

The address-propagation behavior also differs in a way that is directly exploitable. Hidden service nodes exchange onion addresses with incoming Monero peers over internal Tor, but their distinctive self-advertisement appears in Timed Sync Responses sent to outgoing Monero Tor hidden service peers: these responses include up to 249 previously unshared whitelisted onion addresses followed by the node's \emph{own} onion address. By contrast, Timed Sync Responses from Tor client nodes contain only unshared onion addresses. As shown in Figure~\ref{img_onion_analysis}, this makes the last onion address stable across repeated responses from hidden service nodes but variable across responses from Tor client nodes. This distinction forms the basis of the onion address analysis in Section~\ref{subsec-role-identi}, where address repetition across responses is used to identify incoming hidden service peers and retrieve their onion addresses.

\section[Our Approach: ProxyMark]{Our Approach: {\normalfont\itshape ProxyMark}}
\label{sec-approach}

De-anonymizing Monero transactions over Tor requires resolving three uncertainties: what role a Tor-based Monero peer plays, whether a transaction received through a Tor connection is originated by that peer or merely relayed, and which real IP address corresponds to the originating node. \textit{ProxyMark} addresses these uncertainties as a unified pipeline rather than as independent attacks. This section first presents the threat model and design rationale of the pipeline, and then details the operational procedures of its three stages.

\subsection{Threat Model}
\label{subsec-threat-model}

The adversary aims to deanonymize Monero transactions initiated over the Tor network by associating transactions with the IP addresses of their originating nodes. The targets are Monero nodes that comply with the Monero protocol and operate over Tor, including both Monero Tor hidden service nodes and Monero Tor client nodes. The adversary is active at the Monero P2P layer: it can operate multiple Monero Tor hidden service nodes and Monero Tor client nodes, initiate Monero connections to reachable hidden service nodes, exchange protocol messages with target nodes, and run modified adversary-controlled Monero nodes that advertise selected peer addresses or block-height values.

The adversary also operates at the Tor layer by deploying Tor relay nodes. When one of these relays is selected as the entry relay of a target circuit, the adversary can observe circuit-level timing and the source IP address of the target, but cannot decrypt Tor cells or inspect Monero messages carried inside encrypted Tor traffic. To increase the chance that targets connect to adversary-controlled Monero peers, the adversary controls 5,000 online Monero Tor hidden service onion addresses for graylist population. However, the adversary does not compromise target machines, Tor relays outside its control, Tor directory authorities, Monero consensus, or any cryptographic mechanisms used by Tor or Monero.

\subsection{Design Overview}
\label{subsec-approach-overview}

\begin{figure*}[t]
  \centering
  \includegraphics[width=\linewidth]{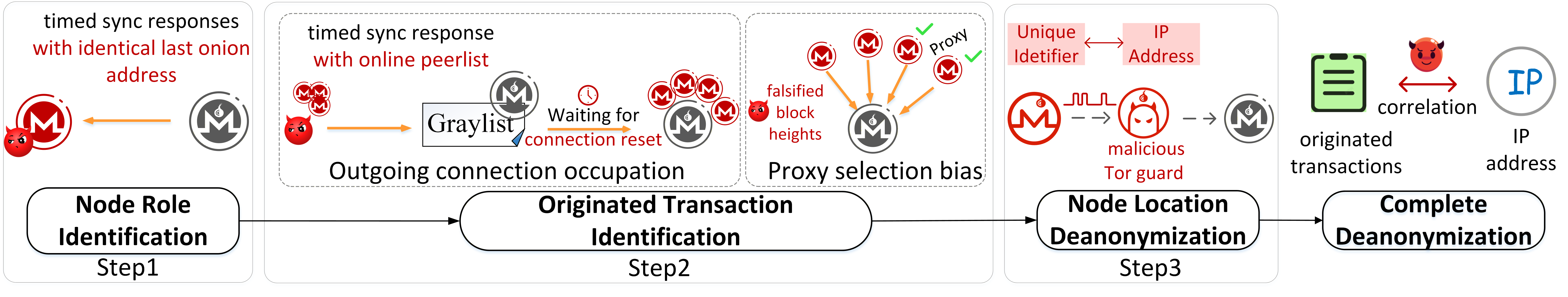}
  \caption{Overall process of \textit{ProxyMark}.}
  \label{img_hs_total}
\end{figure*}
Figure~\ref{img_hs_total} summarizes \textit{ProxyMark}. The framework is organized as a three-stage pipeline: it first identifies whether a Tor peer is a client node or hidden service node, then captures originated transactions by occupying hidden-service outbound connections and biasing proxy selection, and finally links node identifiers to real IP addresses through role-specific watermarking and denoising.

\smallskip
\noindent\textbf{Workflow.} Algorithm~\ref{alg:proxymark_workflow} records how the three stages compose. \textit{ProxyMark} first classifies the peer (line~\ref{algline:role}), assigns a role-specific identifier (lines~\ref{algline:hs-id} and~\ref{algline:client-id}), captures originated transactions through Monero P2P behavior (lines~\ref{algline:hs-capture} and~\ref{algline:client-capture}), and finally binds that identifier to a source IP through watermark detection at an adversarial entry relay (lines~\ref{algline:relay-detect}--\ref{algline:output}). The hidden-service branch uses direct peerlist filling and active, handshake-free watermark injection (lines~\ref{algline:hs-peerlist}--\ref{algline:hs-watermark}), while the client branch uses indirect peerlist population, proxy-selection bias, and noise-cancelled watermarking after an outbound client connection appears (lines~\ref{algline:client-peerlist}--\ref{algline:client-watermark}).

\begin{algorithm}[!b]
\footnotesize
\caption{End-to-end \textit{ProxyMark} workflow}
\label{alg:proxymark_workflow}
\KwIn{Candidate Tor connections $\mathcal{P}$ to adversarial Monero hidden-service peers; adversarial Tor relays $\mathcal{R}$}
\KwData{Role map $L$, identifier map $I$, captured transactions $X$, identifier--IP map $G$}
\ForEach{candidate connection $p \in \mathcal{P}$}{
    $(role,onion) \leftarrow \pmfn{identify\_role}(p)$\nllabel{algline:role}\tcp*[r]{Timed Sync probing}
    $L[p] \leftarrow role$\;
    \uIf{$role = \mathrm{hidden\ service}$}{
        $I[p] \leftarrow \pmfn{encode\_onion\_index}(onion)$\nllabel{algline:hs-id}\tcp*[r]{onion-derived identifier}
        \pmfn{fill\_hidden\_service\_peerlist}$(onion)$\nllabel{algline:hs-peerlist}\tcp*[r]{occupy outgoing peers}
        $X[p] \leftarrow \pmfn{capture\_originated\_txs}(p)$\nllabel{algline:hs-capture}\tcp*[r]{incoming internal-Tor txs}
        \pmfn{inject\_hs\_watermark}$(onion, I[p])$\nllabel{algline:hs-watermark}\tcp*[r]{handshake-free embedding}
    }
    \Else{
        $I[p] \leftarrow \pmfn{assign\_client\_identifier}(p)$\nllabel{algline:client-id}\tcp*[r]{connection-specific identifier}
        \pmfn{populate\_client\_reachable\_peerlists}$(p)$\nllabel{algline:client-peerlist}\tcp*[r]{indirect peerlist filling}
        \pmfn{bias\_proxy\_selection}$(p)$\nllabel{algline:client-bias}\tcp*[r]{height-based proxy bias}
        $X[p] \leftarrow \pmfn{capture\_originated\_txs}(p)$\nllabel{algline:client-capture}\tcp*[r]{captured at malicious peer}
        \pmfn{embed\_client\_watermark}$(p, I[p])$\nllabel{algline:client-watermark}\tcp*[r]{noise cancellation}
    }
    \ForEach{relay $r \in \mathcal{R}$}{
        \If{$r$ detects identifier $I[p]$ on an entry circuit\nllabel{algline:relay-detect}}{
            $G[I[p]] \leftarrow \pmfn{source\_ip}(r, I[p])$\nllabel{algline:source-ip}\tcp*[r]{relay observes source IP}
            \pmfn{output}$(X[p], I[p], G[I[p]])$\nllabel{algline:output}\tcp*[r]{transaction--identifier--IP tuple}
        }
    }
}
\end{algorithm}

\smallskip
\noindent\textbf{Design rationale.} The design of \textit{ProxyMark} is driven by three observations that expose complementary leakage points across the Monero P2P layer and the Tor transport layer.

\smallskip
\noindent\textbf{Observation 1: Role-dependent peer exchange.} Monero Tor hidden service nodes and client nodes exhibit distinct behaviors in Timed Sync Response messages sent over hidden service outbound connections. Specifically, for a Monero Tor hidden service node, the last onion address in each periodic Timed Sync Response message is always identical and corresponds to its own onion address. In contrast, a Monero Tor client node only includes previously unseen onion addresses in such messages; therefore, the last onion address differs across successive Timed Sync Responses. \textit{Method.} \textit{ProxyMark} uses this difference to identify the role of an incoming Tor peer and extract hidden-service onion addresses for later stages.

\smallskip
\noindent\textbf{Observation 2: Proxy-only forwarding of originated transactions.} During each epoch (a randomized interval of 5 to 5.5 minutes), both Monero Tor hidden service nodes and Monero Tor client nodes forward their originated transactions only to two randomly selected hidden service outbound connections whose block heights are higher than their local block height, which we refer to as proxy nodes, while relayed transactions are forwarded to ordinary neighbors. \textit{Method.} \textit{ProxyMark} increases the probability that adversary-controlled hidden service nodes are selected as proxy nodes through outgoing connection occupation and proxy selection bias; transactions received from the corresponding incoming Tor peers are then identified as originated transactions.

\smallskip
\noindent\textbf{Observation 3: Split visibility across Monero and Tor.} A Monero-layer adversary can learn or assign a node identifier but cannot observe the target's real IP address, while a malicious Tor entry relay can observe the target's source IP address but cannot see Monero-level identifiers inside encrypted Tor traffic. \textit{Method.} \textit{ProxyMark} bridges these two partial views by embedding the Monero-level identifier into the timing pattern of Tor relay cells and detecting the encoded identifier at malicious entry relays.

\smallskip
\noindent The following subsections give the operational details of the three stages: node role identification, originated transaction identification, and node location deanonymization.

\subsection{(Step-\ding{182}) Monero Tor Node Role Identification} 
\label{subsec-role-identi}

\smallskip
\noindent\textbf{Probing setup.} The adversary operates a modified Monero Tor hidden service node and treats each incoming Tor connection as a candidate peer whose role must be determined. For each candidate connection, the adversary actively sends Timed Sync Request messages and records the peer lists contained in the corresponding Timed Sync Response messages. This probing uses standard Monero P2P messages over an already established Tor connection and does not require observing or modifying Tor-layer traffic.

\smallskip
\noindent\textbf{Classification rule.} The adversary inspects the last onion address in three consecutive Timed Sync Response messages received from the same incoming peer. If this last address is identical across all three responses, the peer is classified as a Monero Tor hidden service node, and the repeated address is recorded as the peer's onion address. If the last address changes across responses, the peer is classified as a Monero Tor client node. When fewer than three valid responses are available, the adversary continues probing rather than making a role decision from incomplete evidence. This repeated-address rule is the conservative procedure used in our implementation and evaluation; Appendix~\ref{appendix_onion_address_analysis_single} discusses a single-response variant based on the \texttt{last\_seen} timestamp.

\smallskip
\noindent\textbf{Stage output.} The output of this stage is a role label for each incoming Tor peer and, for hidden service nodes, the corresponding onion address. This information determines which strategy is used in later stages: hidden-service onion addresses enable direct peerlist manipulation and serve as natural identifiers, whereas client nodes require indirect peerlist population and client-specific identifier assignment.

\subsection{(Step-\ding{183}) Originated Transaction Identification} 
\label{subsec-originated-tx-identi}

\smallskip
\noindent\textbf{Operational detail.} The adversary deploys multiple Monero Tor hidden service nodes as candidate proxy nodes and waits for benign Monero Tor nodes to establish hidden-service outbound connections with them. Once a malicious hidden service node receives a transaction from an incoming Tor connection selected as a proxy connection, the adversary attributes this transaction to the corresponding incoming peer.

In practice, a large number of benign hidden service nodes already exist in the Monero network. If the adversary deploys only a small number of hidden service nodes, target Monero Tor nodes are likely to connect to benign hidden service nodes and forward their originated transactions accordingly, resulting in a low capture rate for the adversary. To increase capture probability, we use two complementary techniques. The first is \textit{outgoing connection occupation}: the adversary populates target peer lists with malicious onion addresses, making target nodes more likely to establish outbound hidden-service connections to adversary nodes. The second is \textit{proxy selection bias}: after occupying some outbound connections, the adversary advertises fresher block heights so that malicious peers are more likely to pass Monero's proxy selection filter.

\smallskip
\noindent\textbf{Outgoing connection occupation.} By analyzing the outbound connection establishment process of Monero Tor nodes, we observe that outbound connections may be dropped due to node restarts or upon reaching the maximum number of outbound peers, after which new outbound connections are reselected from the graylist and whitelist. Leveraging this mechanism, the adversary aims to populate these lists in advance so that malicious nodes are selected with high probability when outbound connections are reset, thereby enabling outbound connection occupation. The core challenge lies in filling the target node’s graylist and whitelist.

\smallskip
\noindent\textbf{Peerlist filling strategy.} In Monero Tor nodes, the graylist is populated with onion addresses received from other peers, while the whitelist is formed by promoting addresses from the graylist through outbound connections or housekeeping procedures. As a result, the adversary can only directly populate the graylist and must wait for malicious addresses to be gradually promoted to the whitelist. Using the node role identification method in Section~\ref{subsec-role-identi}, the adversary obtains the onion addresses of Monero Tor hidden service nodes and directly injects malicious onion addresses by establishing outbound connections to these nodes, enabling direct graylist population. In contrast, since adversary nodes cannot directly establish outbound connections with Monero Tor client nodes, graylist population for client nodes can only be achieved indirectly.

For Monero Tor hidden service nodes, the adversary directly fills the graylist by sending Timed Sync Responses containing malicious onion addresses. Because each response can carry up to 250 onion addresses and the graylist stores 5,000 entries with first-in-first-out eviction, 20 coordinated malicious nodes are sufficient to fill the target graylist. The adversary delays these responses while staying within Monero's 120-second timeout window, so newly inserted malicious addresses persist longer before being overwritten.

For Monero Tor client nodes, direct graylist filling is not possible because adversary nodes cannot initiate inbound connections to clients. Instead, the adversary first fills the graylists and whitelists of Monero Tor hidden service nodes; these hidden service nodes subsequently propagate malicious onion addresses to client nodes through normal peer exchange, indirectly populating client graylists and whitelists.

Once the target peer lists have been biased, the adversary waits for outbound connection reselection. Reselection occurs after node restarts or when a maximum number of hidden-service outbound connections is enforced, in which case the node periodically drops a fully synchronized peer and chooses a replacement from its graylist and whitelist. Since malicious addresses have been inserted in advance, the target selects adversary nodes with high probability, allowing broad occupation of the target's hidden-service outbound set.

\smallskip
\noindent\textbf{Proxy selection bias.} Monero Tor nodes synchronize blocks only with ordinary nodes, while no block synchronization is performed with hidden service peers. As a result, the block heights of hidden service outbound connections are updated solely through periodic Timed Sync messages exchanged once per minute. This inherently delayed update introduces up to a one-minute lag in block height perception, causing some hidden service peers to be considered outdated and consequently excluded from proxy nodes selection for originated transaction forwarding. Moreover, Monero Tor nodes do not validate the block height information contained in Timed Sync messages, allowing an adversary to inject falsified block heights without being detected.

\smallskip
\noindent\textbf{Injecting falsified block heights.} To perform block height manipulation, the adversary constructs modified Monero Tor hidden service nodes that actively inject falsified block height information. Specifically, the adversary modifies the Monero node implementation as follows. First, the adversary advertises falsified block heights to target nodes via Timed Sync messages, where the advertised block height is set to the current network height plus a small offset (e.g., $+5$ blocks). Second, the adversary increases the frequency of Timed Sync messages sent to neighbors from the default once per minute to once every 20 seconds, ensuring that the falsified block height information is refreshed more frequently than that of benign peers. In addition, the modified adversary nodes do not synchronize the full blockchain. Instead, they obtain the latest block height from ordinary nodes and use this information solely to construct falsified block height advertisements toward Tor nodes. This design allows the adversary to perform block height manipulation without maintaining a full blockchain replica, significantly reducing storage overhead while remaining effective.

\subsection{(Step-\ding{184}) Node Location Deanonymization} 
\label{subsec-location-deanonymization}

\begin{figure}[pt]
  \centering
  \begin{subfigure}[pt]{0.4\textwidth}
      \includegraphics[width=\linewidth]{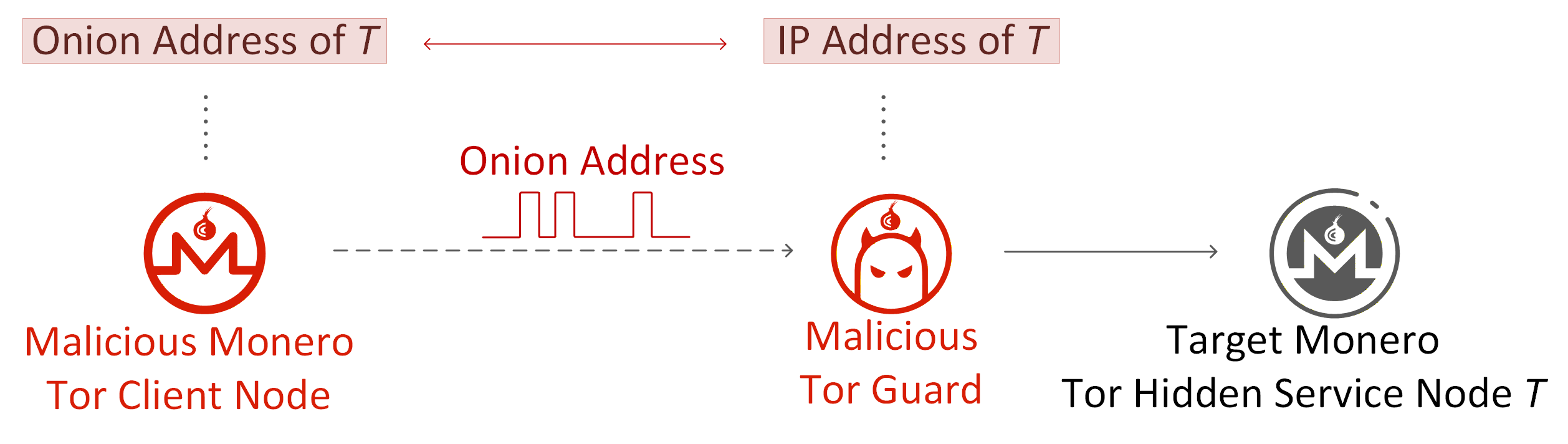}
      \caption{Method to locate a Monero Tor hidden service node}
      \label{img_location_deanonymization_hidden}
  \end{subfigure}
  \begin{subfigure}[pt]{0.4\textwidth}
        \includegraphics[width=\linewidth]{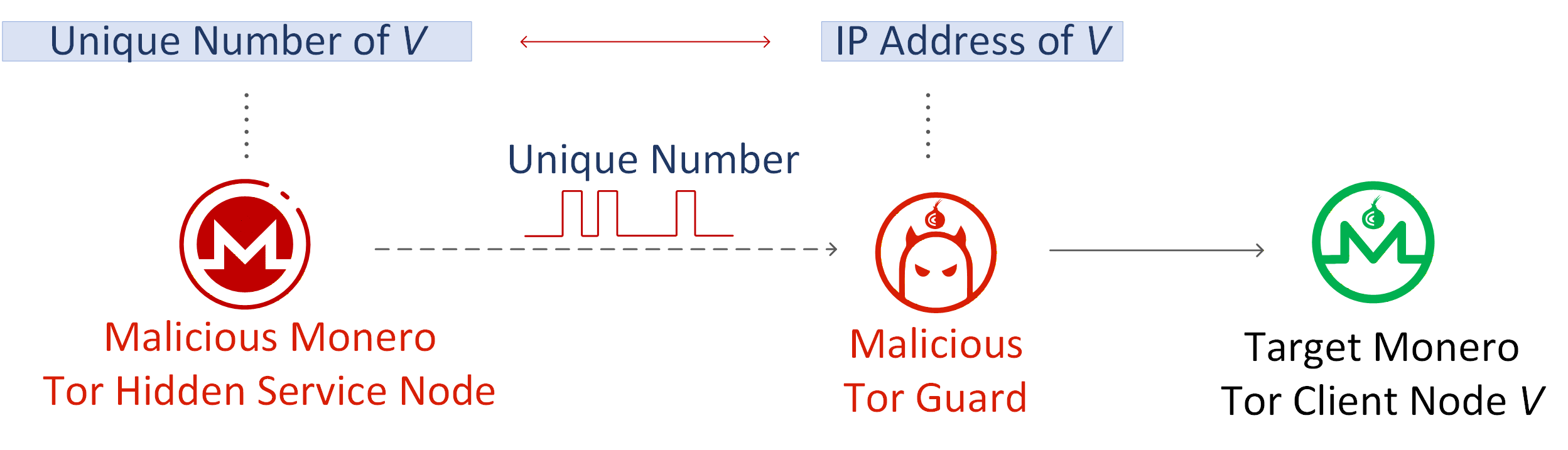}
      \caption{Method to locate a Monero Tor client node}
      \label{img_location_deanonymization_client}
  \end{subfigure}
  \caption{Method to locate a Monero Tor node.}
  \label{img_location_deanonymization}
\end{figure}

\smallskip
\noindent\textbf{Role-specific setup.}
The deployment strategy depends on the target role, as illustrated in Figure~\ref{img_location_deanonymization}. For hidden service nodes, which accept incoming Tor connections, the adversary actively connects to the target and injects the watermark on this adversary-initiated connection. For Tor client nodes, which cannot accept inbound Tor connections, the adversary waits until the client establishes an outbound connection to a malicious hidden service node and then performs watermark embedding on that connection. This setup determines which Monero connection carries the identifier and which Tor circuits are scanned by malicious entry relays.

\smallskip
\noindent\textbf{Identifier assignment.}
For hidden service targets, the semantic identifier is the onion address, encoded for transmission as an assigned index; the 179 active hidden service nodes in our measurement fit in 8 bits. For Tor client targets, the identifier concatenates the malicious hidden-service-node index and the per-connection index. Our 5,000-node/116-connection parameterization therefore uses $13+8=21$ bits.

\smallskip
\noindent\textbf{Watermark embedding.}
The adversary embeds identifiers by controlling the number of Monero signal carrier messages sent in fixed time windows. We use Timed Sync Request messages as carriers because they are Monero application-layer messages, Monero does not rate-limit them, and unsolicited response messages would cause peers to disconnect. Ping and Support Flags requests are also viable carriers, but our implementation uses Timed Sync Requests. To reduce message aggregation at the Tor client, carrier messages sent within the same window are separated by at least 0.2 seconds; we use a 6-second window in our implementation.

The first two windows form an initialization pattern that marks the watermark-carrying circuit. The adversary sends $n$ carrier messages in each initialization window, with $n=5$ for hidden service nodes and $n=6$ for client nodes. The remaining windows encode the identifier bit by bit: three carrier messages represent binary 1, and one carrier message represents binary 0. For example, a hidden service identifier encoded as \texttt{10010110} yields carrier counts $\langle 5,5,3,1,1,3,1,3,3,1\rangle$, where the first two entries are the initialization pattern.

\smallskip
\noindent\textbf{Signal detection.}
The malicious Tor relay first determines whether it is the entry relay of a candidate circuit. This can be inferred from circuit-establishment traffic: for hidden service nodes, the entry relay receives two \texttt{RELAY} cells directed to the hidden service, while an intermediate relay receives one; for Tor client nodes, the entry relay receives one \texttt{RELAY} cell directed to the client, while an intermediate relay receives none. The initialization pattern separates circuit-establishment traffic from application-layer watermark traffic.

After filtering candidate entry circuits, the malicious relay scans inbound \texttt{RELAY} cells for the watermark pattern. It treats each second as a possible start time, counts cells across subsequent time windows, checks for the two initialization windows, and then decodes each following window as either 1 or 0 according to the observed carrier count. Once a valid identifier is decoded, the relay associates it with the observed source IP address of the circuit.

\smallskip
\noindent\textbf{Noise handling.}
Watermark detection must tolerate protocol traffic that is not part of the intended signal. The main noise sources are Monero white-noise traffic, Timed Sync Responses required by the Monero protocol, and Tor flow-control cells such as \texttt{RELAY\_SENDME}. Because Tor cells are encrypted, a malicious relay cannot distinguish these noise cells from carrier cells by content.

For hidden service targets, we avoid most protocol-induced noise through handshake-free watermark embedding. Monero permits a peer to transmit up to 256~KB before completing the handshake, and the selected Timed Sync Request carriers fit within this limit. Therefore, a malicious Monero Tor client can send the watermark without completing the Monero handshake, avoiding white-noise traffic and Timed Sync Response obligations. This also allows watermark injection even when the target hidden service has reached its incoming-connection limit, because no full handshake request is made.

For Tor client targets, handshake-free embedding is not applicable because the adversary must wait for the client to connect to a malicious hidden service node. We therefore use noise cancellation: protocol-required noise messages are deferred to the next signal window and used to replace carrier messages whenever possible. Aligning unavoidable noise with the expected signal pattern prevents it from corrupting the decoded identifier.

\section{Experiments}
\label{sec-experi}

This section reports component-wise experiments for the three stages of \textit{ProxyMark}: role identification, originated-transaction identification, and node-location deanonymization.

\subsection{Experimental Settings}
\label{subsec:eval_settings}

We evaluate the three stages of \textit{ProxyMark} separately: role identification, originated-transaction identification, and node-location deanonymization. Experiment-specific parameters, resource scaling, and ethical constraints are reported with the corresponding experiment rather than centralized here.

\smallskip
\noindent\textbf{Environments.} We evaluated role identification on the Monero testnet, proxy selection bias on the Monero mainnet, connection occupation in a controlled live Tor hidden-service deployment, and watermarking with Monero testnet nodes over the live Tor network. We used Monero v0.18.3.1 target nodes in the role-identification, proxy-bias, and watermarking experiments. We ran Tor relays with Tor v0.4.7.13 for hidden-service location experiments and Tor v0.4.8.10 for client-node location experiments.

\smallskip
\noindent\textbf{Metrics.} For role identification and location deanonymization, we report precision and recall. For connection occupation, we report the number of adversarial addresses in the target's outgoing connections, graylist, and whitelist. For proxy selection bias, we report the baseline proxy-selection rate, the biased proxy-selection rate, the gain ratio, and the absolute gain.

\subsection{Evaluation of Monero Tor Node Role Identification}

We evaluate Step-\ding{182} by testing whether incoming Tor peer-list behavior is sufficient to distinguish hidden-service nodes from Tor client nodes and recover the onion address of each hidden-service node.

\noindent\textbf{Setup and design.} We modified the Monero node source code to implement the onion address analysis method. The modified hidden service node examines the last onion address in three consecutive Timed Sync Response messages received from each incoming Tor connection. If the address is identical across all three responses, the incoming connection is classified as a hidden service node and its onion address is recorded; otherwise, it is classified as a Tor client node.

To validate the effectiveness of the method, we deployed a four-node testnet environment on the Monero testnet: one adversarial hidden service node $M$ running the modified software, two target hidden service nodes $A$ and $B$, and one target Tor client node $C$. All three target nodes ran Monero v0.18.3.1. Nodes $A$, $B$, and $C$ each established 100 independent connections to node $M$, for a total of 300 connections. Node $M$ applied the onion address analysis method to associate each incoming connection with the corresponding node role and onion address.

\smallskip
\noindent\textbf{Results.}
For all 200 connections from nodes $A$ and $B$, node $M$ successfully associated each connection with the correct onion address. For all 100 connections from node $C$, node $M$ correctly identified them as Tor client connections and assigned no onion address. The method achieves 100\% precision and 100\% recall, confirming that role-dependent peer-list behavior is stable enough to support reliable role classification and hidden-service onion-address extraction. The deterministic nature of this result is expected, as the identification method is grounded in deterministic protocol reasoning rather than probabilistic inference.

\subsection{Evaluation of Originated Transaction Identification}

We evaluate Step-\ding{183} by testing two conditions for capturing originated transactions: whether peer-list filling can increase the adversary's share of hidden-service outbound connections, and whether occupied adversarial peers can bias proxy choice.

\subsubsection[Connection Occupation]{\underhead{Connection Occupation}}
\label{subsubsec:eval_occupation}

This experiment tests whether graylist and whitelist filling can translate adversarial onion addresses into outbound hidden-service connections under two reset conditions: node restart and periodic replacement after the maximum outbound limit is reached. It instantiates the 5,000-address adversary in the threat model at a 10$\times$ scaled size because of resource and ethical constraints. We constructed 500 Monero hidden service identities by hosting multiple Tor hidden services on the same Tor process and mapping them to shared Monero backends. A single Tor process can host up to approximately 25 Tor hidden services. We deployed multiple Tor processes per machine, setting $N_{\text{Tor}} = N_{\text{CPU}} - 2$ to reserve two CPU cores for the Monero node and system services. The cluster comprised two dedicated 8-core CPU servers, each running 6 Tor processes ($6 \times 25 = 150$ hidden service nodes per server), and two shared 48-core CPU servers, each running 4 Tor processes ($4 \times 25 = 100$ hidden service nodes per server). On the shared servers, the process count was capped at 4 because higher counts caused instability due to resource contention.

To preserve the scaled ratio, the target node's graylist capacity was reduced from 5,000 to 500 and its whitelist capacity from 1,000 to 100. Addresses received from adversary nodes were stored at the normal rate of 250 addresses per Timed Sync Response, while only 25 randomly selected addresses from benign nodes were stored. For ethical reasons, adversary nodes did not forward onion addresses to their neighbors, avoiding contamination of the Monero network.

We ran a target hidden service node on January 25, 2025. A Python graylist-filling program connected two adversary Monero nodes to the target; upon each Timed Sync Request, the adversary returned 250 adversarial onion addresses after a 5-second delay. For measurement only, the last address in each adversary response was set to the sentinel address \texttt{55...55.onion}.

\smallskip
\noindent\textbf{Node restart.} Because a restarted node selects outbound connections from both the whitelist and graylist, graylist filling was completed before each restart. The graylist filling phase took approximately 1 minute, and whitelist filling took approximately 2 hours (estimated at $\sim$20 hours for a full-capacity whitelist of 1{,}000 entries). Once the whitelist is fully occupied, adversarial addresses are no longer reinserted into the graylist, so fully saturating the graylist after complete whitelist occupation may require additional resources.

Across three independent repetitions, we repeatedly restarted the target while maintaining graylist filling. Figure~\ref{img_restart} reports the adversarial addresses in outbound connections, graylist, and whitelist. After one or two restarts, the adversary stably occupied 7 to 11 of the 12 hidden service outbound connections. The increasing trend follows Monero's outbound-selection rule: 8 slots are preferentially selected from recent whitelist entries and 4 slots are drawn from the graylist; once adversarial peers enter the outbound set, their \texttt{last\_seen} timestamps are refreshed, making later selection more likely.

\definecolor{pmRed}{RGB}{218,42,42}
\definecolor{pmGreen}{RGB}{43,139,87}
\definecolor{pmGold}{RGB}{224,163,20}

\newcommand{\pmRestartLegend}{%
  \begin{tikzpicture}[baseline=-0.55ex]
    \draw[pmRed, line width=1.15pt] (0,0) -- (0.48,0);
    \node[circle, draw=pmRed, line width=0.3pt, inner sep=0.85pt] at (0.24,0) {};
    \node[anchor=west, font=\footnotesize] at (0.56,0) {Whitelist};
    \draw[pmGreen, dashed, dash pattern=on 4pt off 2pt, line width=1.15pt] (2.28,0) -- (2.76,0);
    \node[regular polygon, regular polygon sides=4, draw=pmGreen, line width=0.3pt, inner sep=0.85pt] at (2.52,0) {};
    \node[anchor=west, font=\footnotesize] at (2.84,0) {Graylist};
    \draw[pmGold, line width=1.15pt] (4.36,0) -- (4.84,0);
    \node[regular polygon, regular polygon sides=3, draw=pmGold, line width=0.3pt, inner sep=1pt, rotate=0] at (4.60,0) {};
    \node[anchor=west, font=\footnotesize] at (4.92,0) {Outgoing};
  \end{tikzpicture}%
}

\pgfplotsset{
  pmRestartAxis/.style={
    width=\linewidth,
    height=0.72\linewidth,
    xmin=-0.35, xmax=6.35,
    ymin=-18, ymax=520,
    axis lines=box,
    axis line style={draw=black, line width=0.5pt},
    tick align=outside,
    tick style={draw=black, line width=0.4pt},
    grid=major,
    major grid style={draw=gray!24, line width=0.25pt},
    xtick={0,1,2,3,4,5,6},
    xticklabels={{\shortstack{After\\Addr. fill}},1,2,3,4,5,6},
    ytick={0,100,200,300,400,500},
    xlabel={Restart count},
    label style={font=\footnotesize},
    tick label style={font=\scriptsize},
    clip=false,
  },
  pmRedPlot/.style={
    pmRed,
    line width=1.05pt,
    mark=o,
    mark size=0.95pt,
    mark options={solid, draw=pmRed, fill opacity=0, line width=0.3pt},
  },
  pmGreenPlot/.style={
    pmGreen,
    dashed,
    dash pattern=on 4pt off 2pt,
    line width=1.05pt,
    mark=square,
    mark size=0.9pt,
    mark options={solid, draw=pmGreen, fill opacity=0, line width=0.3pt},
  },
  pmGoldPlot/.style={
    pmGold,
    line width=1.05pt,
    mark=triangle,
    mark size=1.1pt,
    mark options={solid, draw=pmGold, fill opacity=0, line width=0.3pt},
  },
}

\begin{figure*}[t]
  \centering
  \pmRestartLegend
  \par\vspace{3pt}

  \begin{subfigure}[t]{0.32\textwidth}
    \centering
    \begin{tikzpicture}
      \begin{axis}[pmRestartAxis, ylabel={No. of attacker addr.}]
        \addplot+[pmRedPlot] coordinates {(0,88) (1,82) (2,81) (3,79) (4,79) (5,78) (6,74)};
        \addplot+[pmGreenPlot] coordinates {(0,411) (1,387) (2,387) (3,300) (4,416) (5,311) (6,422)};
        \addplot+[pmGoldPlot, point meta=explicit symbolic, nodes near coords,
          every node near coord/.append style={font=\fontsize{5}{5}\selectfont\bfseries, text=pmGold, fill=white, inner sep=0.25pt, yshift=2pt}]
          coordinates {(0,1) [1] (1,4) [4] (2,7) [7] (3,9) [9] (4,10) [10] (5,8) [8] (6,8) [8]};
      \end{axis}
    \end{tikzpicture}
    \caption{First repeated experiment}
    \label{img_restart1}
  \end{subfigure}
  \hfill
  \begin{subfigure}[t]{0.32\textwidth}
    \centering
    \begin{tikzpicture}
      \begin{axis}[pmRestartAxis, yticklabels=\empty]
        \addplot+[pmRedPlot] coordinates {(0,88) (1,83) (2,82) (3,81) (4,79) (5,79) (6,79)};
        \addplot+[pmGreenPlot] coordinates {(0,410) (1,354) (2,407) (3,399) (4,267) (5,392) (6,333)};
        \addplot+[pmGoldPlot, point meta=explicit symbolic, nodes near coords,
          every node near coord/.append style={font=\fontsize{5}{5}\selectfont\bfseries, text=pmGold, fill=white, inner sep=0.25pt, yshift=2pt}]
          coordinates {(0,1) [1] (1,7) [7] (2,7) [7] (3,10) [10] (4,8) [8] (5,9) [9] (6,10) [10]};
      \end{axis}
    \end{tikzpicture}
    \caption{Second repeated experiment}
    \label{img_restart2}
  \end{subfigure}
  \hfill
  \begin{subfigure}[t]{0.32\textwidth}
    \centering
    \begin{tikzpicture}
      \begin{axis}[pmRestartAxis, yticklabels=\empty]
        \addplot+[pmRedPlot] coordinates {(0,88) (1,80) (2,80) (3,77) (4,76) (5,76) (6,75)};
        \addplot+[pmGreenPlot] coordinates {(0,412) (1,418) (2,334) (3,323) (4,420) (5,373) (6,309)};
        \addplot+[pmGoldPlot, point meta=explicit symbolic, nodes near coords,
          every node near coord/.append style={font=\fontsize{5}{5}\selectfont\bfseries, text=pmGold, fill=white, inner sep=0.25pt, yshift=2pt}]
          coordinates {(0,1) [1] (1,7) [7] (2,10) [10] (3,10) [10] (4,9) [9] (5,11) [11] (6,8) [8]};
      \end{axis}
    \end{tikzpicture}
    \caption{Third repeated experiment}
    \label{img_restart3}
  \end{subfigure}

  \caption{Experimental results of outgoing connection occupation after adversarial address filling.}
  \label{img_restart}
\end{figure*}
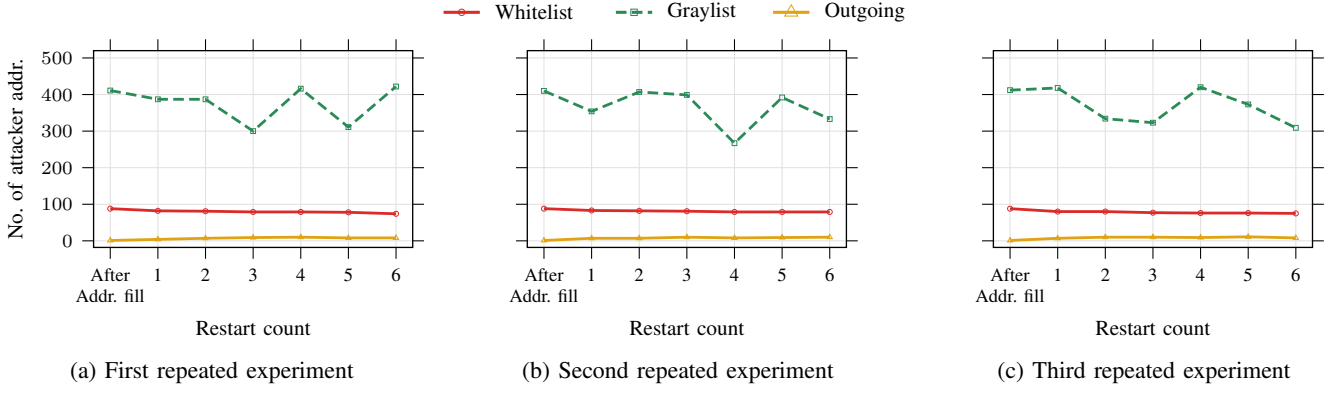

\smallskip
\noindent\textbf{Maximum outbound connection limit.} In this reset scenario, the target drops one fully synchronized hidden service outbound connection every 101 seconds and replaces it with a new connection drawn from the graylist. Thus, graylist filling alone is sufficient. We set the target's maximum number of hidden service outbound connections to 10, ran three repetitions, and recorded adversarial addresses in outbound connections, graylist, and whitelist every 3 seconds after the initial connections formed.

Figure~\ref{img_update} shows that adversarial addresses quickly saturated the graylist and then progressively occupied the outbound set. Across all three repetitions, approximately 20 minutes after the attack commenced, the adversary stably occupied 8 to 10 of the 10 outbound connections.

\definecolor{pmRed}{RGB}{218,42,42}
\definecolor{pmGreen}{RGB}{43,139,87}
\definecolor{pmGold}{RGB}{224,163,20}

\newcommand{\pmUpdateLegend}{%
  \begin{tikzpicture}[baseline=-0.55ex]
    \draw[pmRed, line width=1.15pt] (0,0) -- (0.48,0);
    \node[circle, fill=pmRed, inner sep=1.15pt] at (0.24,0) {};
    \node[anchor=west, font=\footnotesize] at (0.56,0) {Whitelist};
    \draw[pmGreen, dashed, dash pattern=on 4pt off 2pt, line width=1.15pt] (2.28,0) -- (2.76,0);
    \node[regular polygon, regular polygon sides=4, fill=pmGreen, inner sep=1.15pt] at (2.52,0) {};
    \node[anchor=west, font=\footnotesize] at (2.84,0) {Graylist};
    \draw[pmGold, line width=1.15pt] (4.36,0) -- (4.84,0);
    \node[regular polygon, regular polygon sides=3, fill=pmGold, inner sep=1.35pt, rotate=0] at (4.60,0) {};
    \node[anchor=west, font=\footnotesize] at (4.92,0) {Outgoing};
  \end{tikzpicture}%
}

\pgfplotsset{
  pmUpdateAxis/.style={
    width=\linewidth,
    height=0.70\linewidth,
    xmin=-1, xmax=69,
    ymin=-18, ymax=520,
    axis lines=box,
    axis line style={draw=black, line width=0.5pt},
    tick align=outside,
    tick style={draw=black, line width=0.4pt},
    grid=major,
    major grid style={draw=gray!24, line width=0.25pt},
    xtick={0,10,20,30,40,50,60,68},
    ytick={0,100,200,300,400,500},
    xlabel={Relative time (min)},
    label style={font=\footnotesize},
    tick label style={font=\scriptsize},
    clip=false,
  },
  pmRedPlot/.style={
    pmRed,
    line width=1pt,
    mark=*,
    mark size=1.15pt,
    mark options={solid, fill=pmRed},
  },
  pmGreenPlot/.style={
    pmGreen,
    dashed,
    dash pattern=on 4pt off 2pt,
    line width=1.15pt,
    mark=square*,
    mark size=1.15pt,
    mark options={solid, fill=pmGreen},
  },
  pmGoldPlot/.style={
    pmGold,
    line width=1.15pt,
    mark=triangle*,
    mark size=1.05pt,
    mark options={solid, fill=pmGold},
  },
}

\begin{figure*}[t]
  \centering
  \pmUpdateLegend
  \par\vspace{3pt}

  \begin{subfigure}[t]{0.32\textwidth}
    \centering
    \begin{tikzpicture}
      \begin{axis}[pmUpdateAxis, ylabel={No. of attacker addr.}]
        \addplot+[pmRedPlot] coordinates {(0,0) (4,5) (7,7) (11,12) (14,16) (18,19) (21,22) (25,25) (28,30) (32,33) (36,38) (39,43) (43,45) (46,48) (50,54) (53,57) (57,61) (60,66) (64,70) (68,75)};
        \addplot+[pmGreenPlot] coordinates {(0,0) (4,490) (7,489) (11,335) (14,384) (18,440) (21,471) (25,467) (28,455) (32,458) (36,453) (39,448) (43,439) (46,440) (50,411) (53,431) (57,425) (60,420) (64,416) (68,403)};
        \addplot+[pmGoldPlot] coordinates {(0,0) (4,1) (7,2) (11,4) (14,7) (18,8) (21,9) (25,10) (28,10) (32,10) (36,10) (39,10) (43,9) (46,8) (50,9) (53,9) (57,8) (60,8) (64,9) (68,9)};
      \end{axis}
    \end{tikzpicture}
    \caption{First repeated experiment}
    \label{img_update1}
  \end{subfigure}
  \hfill
  \begin{subfigure}[t]{0.32\textwidth}
    \centering
    \begin{tikzpicture}
      \begin{axis}[pmUpdateAxis, yticklabels=\empty]
        \addplot+[pmRedPlot] coordinates {(0,0) (4,5) (7,8) (11,13) (14,17) (18,19) (21,23) (25,29) (29,35) (32,40) (36,45) (39,50) (43,54) (46,59) (50,61) (53,65) (57,71) (61,76) (64,81) (68,85)};
        \addplot+[pmGreenPlot] coordinates {(0,0) (4,251) (7,492) (11,459) (14,481) (18,477) (21,472) (25,466) (29,461) (32,441) (36,450) (39,446) (43,440) (46,436) (50,433) (53,428) (57,422) (61,417) (64,413) (68,411)};
        \addplot+[pmGoldPlot] coordinates {(0,0) (4,1) (7,2) (11,5) (14,7) (18,8) (21,9) (25,10) (29,10) (32,10) (36,10) (39,10) (43,10) (46,10) (50,10) (53,10) (57,10) (61,10) (64,10) (68,9)};
      \end{axis}
    \end{tikzpicture}
    \caption{Second repeated experiment}
    \label{img_update2}
  \end{subfigure}
  \hfill
  \begin{subfigure}[t]{0.32\textwidth}
    \centering
    \begin{tikzpicture}
      \begin{axis}[pmUpdateAxis, yticklabels=\empty]
        \addplot+[pmRedPlot] coordinates {(0,0) (4,5) (7,8) (11,13) (14,17) (18,22) (21,25) (25,31) (28,36) (32,40) (36,45) (39,47) (43,52) (46,57) (50,63) (53,68) (57,71) (61,75) (64,79) (68,85)};
        \addplot+[pmGreenPlot] coordinates {(0,0) (4,382) (7,495) (11,489) (14,459) (18,479) (21,476) (25,470) (28,457) (32,460) (36,456) (39,451) (43,444) (46,440) (50,435) (53,430) (57,425) (61,418) (64,412) (68,409)};
        \addplot+[pmGoldPlot] coordinates {(0,0) (4,1) (7,3) (11,5) (14,6) (18,7) (21,8) (25,9) (28,9) (32,9) (36,9) (39,9) (43,9) (46,9) (50,9) (53,9) (57,9) (61,10) (64,10) (68,10)};
      \end{axis}
    \end{tikzpicture}
    \caption{Third repeated experiment}
    \label{img_update3}
  \end{subfigure}

  \caption{Experimental results of outgoing connection occupation during periodic connection replacement.}
  \label{img_update}
\end{figure*}
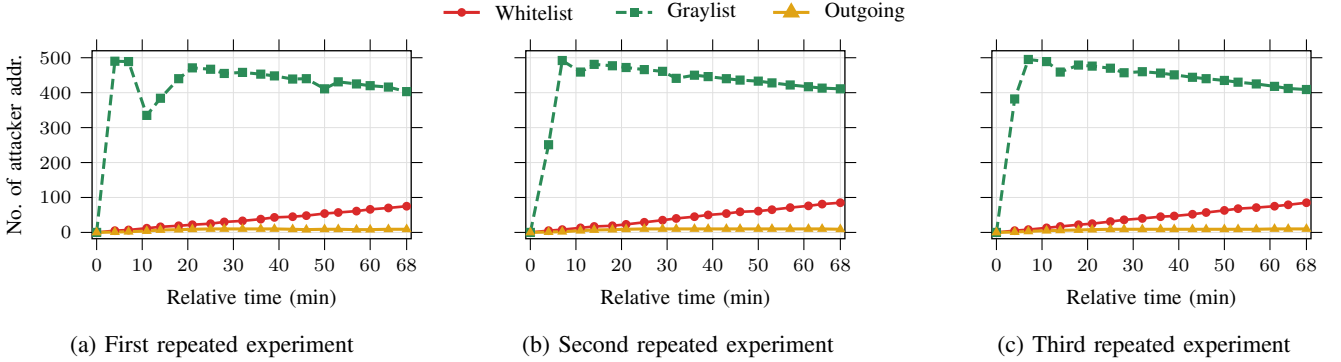

\subsubsection[Proxy Bias]{\underhead{Proxy Bias}}
\label{subsubsec:eval_proxy_bias}
We next measure whether block-height falsification increases the probability that adversary-controlled outbound peers are selected as proxy nodes. The modified nodes report the current network height plus 5, send outgoing Timed Sync Requests every 20 seconds (three times the normal rate), and obtain the latest height from clearnet peers without synchronizing a full blockchain.

Experiments were conducted on the Monero mainnet from April~18 to 25, 2024, with one synchronized Monero Tor client node (v0.18.3.1) as the target. The target maintained 12 hidden service outbound connections; for each experiment, $n \in \{1,3,6,9\}$ of them were adversary-controlled and the remaining $12-n$ were benign. The random-selection probability that at least one adversary node is selected as a proxy is:
\begin{equation}
  p = 1 - \frac{C_{12-n}^{2}}{C_{12}^{2}}
    = 1 - \frac{C_{12-n}^{2}}{66}.
  \label{eq:theory_prob}
\end{equation}

For each $n$, we compare the theoretical probability, the observed baseline rate without proxy bias, the observed bias rate, the gain ratio ($\hat p_{\mathrm{bias}}/\hat p_{\mathrm{base}}$), and the absolute gain ($\hat p_{\mathrm{bias}}-\hat p_{\mathrm{base}}$). Proxy selections per epoch were recorded via node logs; results are shown in Table~\ref{tab:bias_results}.

\begin{table}[tbp]
    \centering
  \caption{Proxy Selection Bias Experimental Results}
  \label{tab:bias_results}
  \footnotesize
  \begin{adjustbox}{max width=\columnwidth}
  \begin{tabular}{cccccr}
    \toprule
    $n$ &
    \makecell{\textbf{Theoretical}\\\textbf{Prob.}} &
    \makecell{\textbf{Baseline}\\\textbf{Rate}} &
    \makecell{\textbf{Bias}\\\textbf{Rate}} &
    \makecell{\textbf{Gain}\\\textbf{Ratio}} &
    \makecell{\textbf{Absolute}\\\textbf{Gain}} \\
    \midrule
    1 & 16.7\% & $24/157=15.3\%$ & $61/171=35.7\%$  & 233\% & $+20.4\%$ \\
    3 & 45.0\% & $49/115=42.6\%$ & $171/277=61.7\%$ & 145\% & $+19.1\%$ \\
    6 & 77.0\% & $83/108=76.9\%$ & $44/52=84.6\%$   & 110\% & $+7.7\%$  \\
    9 & 95.0\% & $56/59=94.9\%$  & $45/45=100\%$    & 105\% & $+5.1\%$  \\
    \bottomrule
  \end{tabular}
  \end{adjustbox}
\end{table}

Proxy bias consistently increases adversarial proxy selection across all tested values of $n$. When $n=1$, the rate increases from 15.3\% to 35.7\% (233\% gain ratio); in 20 of 171 biased selections, all benign outbound hidden service peers appeared stale while the adversary's falsified height exceeded the target's local height, making the adversary the sole eligible proxy candidate. When $n=9$, at least one adversarial peer was selected in all 45 observed biased selections. The baseline rates remain close to the theoretical probabilities (average deviation $<$2\%), and the gain ratio decreases from 233\% to 105\% as adversarial peers increasingly compete with each other. Together, the connection occupation and proxy bias results show that the adversary can first increase its share of the target's outbound hidden-service connections and then further increase the chance that those peers are selected for originated transaction forwarding.

\subsection{Evaluation of Node Location Deanonymization}
\label{subsec:eval_node_locati_deanon}

We evaluate Step-\ding{184} over Monero testnet nodes and the live Tor network, measuring whether the watermarking mechanism can recover role-specific identifiers at an adversarial entry relay and bind them to the observed source IP address. We use a controlled guard configuration: in each experiment, the target node uses the adversarial relay as its sole guard relay, so the relay observes the source IP address for circuits built by the target. This isolates watermark detection and identifier--IP linking from the separate question of adversarial guard placement, which is analyzed in Section~\ref{sec-discuss} and Appendix~\ref{appendix_malicious_guard_selection}. The two target roles are evaluated separately because the identifier enters the traffic at different points.

\smallskip
\noindent\textbf{Hidden-service node targets.} For a hidden-service target, the adversary injects the identifier through an incoming Monero Tor client connection. We deployed a target Monero hidden service node $T$ (v0.18.3.1), an adversarial Monero Tor client node $C$ that sends the watermark signal, and an adversarial Tor relay $R$ (v0.4.7.13) with the \texttt{Guard} flag. Node $T$ used $R$ as its sole guard relay. In each of five independent repetitions, $C$ opened 100 independent Tor circuits to $T$ and encoded identifiers 1 through 100 in binary; $R$ then recorded which identifiers were detected at the entry relay.

\smallskip
\noindent\textbf{Tor client node targets.} The Tor-client case reverses the direction of injection: the target initiates a connection to an adversarial hidden service node, and the adversary embeds the identifier on the response path. We deployed a target Monero Tor client node $T$ (v0.18.3.1), an adversarial Monero hidden service node $H$, and an adversarial Tor relay $R$ (v0.4.8.10) with the \texttt{Guard} flag; again, $T$ used $R$ as its sole guard relay. In each of five independent repetitions, $T$ established 100 connections to $H$, each over an independent Tor circuit. For each connection, $H$ generated a random unique identifier consisting of the adversary node index and a per-connection index, embedded it as a watermark signal, and $R$ recorded the detection result.

\begin{table}[tbp]
  \centering
  \caption{Node Location Deanonymization Results}
  \label{tab:node_loca_deanon}
  \small
  \begin{tabular}{ccccc}
    \toprule
    \multirow{2}{*}{\textbf{Repetition} }
    & \multicolumn{2}{c}{\textbf{Hidden Service Node}} 
    & \multicolumn{2}{c}{\textbf{Tor Client Node}} \\
    \cmidrule(lr){2-3}
    \cmidrule(lr){4-5}
    & \textbf{Precision} & \textbf{Recall}
    & \textbf{Precision} & \textbf{Recall} \\
    \midrule
    1 & 100\% & 94\% & 100\% & 89\% \\
    2 & 100\% & 98\% & 100\% & 93\% \\
    3 & 100\% & 92\% & 100\% & 90\% \\
    4 & 100\% & 88\% & 100\% & 95\% \\
    5 & 100\% & 97\% & 100\% & 90\% \\
    \midrule
    Average & 100\% & 93.8\% & 100\% & 91.4\% \\
    \bottomrule
  \end{tabular}
\end{table}

\smallskip
\noindent Table~\ref{tab:node_loca_deanon} reports precision and recall for both target roles. Precision is the fraction of detected identifiers that match the injected identifiers; recall is the fraction of injected identifiers successfully detected by relay $R$. For hidden-service targets, the method achieves 100\% precision and an average recall of 93.8\%. Failed detections were traced to excessive packet delay in the Tor network and occasional Tor circuit teardown. For Tor client targets, the method achieves 100\% precision and an average recall of 91.4\%. Failed detections were primarily attributed to relay position identification failures and excessive Tor packet delay. Overall, conditional on adversarial entry-relay placement, the watermarking experiments show that \textit{ProxyMark} can link role-specific Monero identifiers to source IP addresses for both target roles.

\smallskip
\noindent\textbf{Practical implication for hidden-service targets.} In the real Tor network, each Tor hidden service selects two guard relays. When the target node's guard set includes the adversarial relay, the adversary can repeatedly establish connections to the target so that at least one watermark-carrying circuit is likely to use the adversarial relay as the circuit entry. When the target selects exactly one adversarial guard relay, the probability that a given circuit uses it as the entry relay is $\tfrac{1}{2}$. The probability that a single watermark injection succeeds is therefore $93.8\% \times \tfrac{1}{2} = 46.9\%$. After $k$ injection attempts, the probability of at least one successful deanonymization is:
\begin{equation}
  P(k) = 1 - (1 - 0.469)^{k}.
  \label{eq:multi_inject}
\end{equation}
With $k=8$, $P(8) = 99.4\%$, enabling the adversary to associate the target node's IP address with its onion address with high probability under this condition.

\section{Discussion and Mitigation}
\label{sec-discuss}

This section summarizes practical considerations and protocol-level defenses for each stage of \textit{ProxyMark}.

\subsection{Practical Considerations and Scope}
The end-to-end attack has three concrete requirements. First, the target must use Monero over Tor and expose the role-specific behavior analyzed in Section~\ref{sec-bck}. Second, the adversary must operate Monero hidden service peers that can interact with the target, fill peer lists, and receive transactions through adversary-controlled outbound connections. Third, the location-deanonymization stage requires at least one adversarial Tor relay to be selected as the target's entry guard. The first two requirements are Monero P2P interactions evaluated in Section~\ref{sec-experi}; the third is the main bottleneck at the Tor layer.

The full derivation of adversarial guard selection is provided in Appendix~\ref{appendix_malicious_guard_selection}; here we summarize the implication. This probability differs for hidden service nodes and Tor client nodes because their circuits are constructed differently. With 179 malicious machines each running one 20 MB/s Tor relay, the probability is $p_e=0.12$ for hidden service nodes and $p_{ec}=0.06$ for Tor client nodes; scaling to 1000 such machines increases these probabilities to $0.46$ and $0.27$, respectively. For hidden service nodes, repeated watermark injections further amplify the success probability once an adversarial relay is in the guard set.

Biryukov's Bitcoin-over-Tor address-cookie attack~\cite{Biryukov2015} does not transfer cleanly to Monero; Appendix~\ref{appendix_feasibilityof_Biryukov_in_Monero} gives the full comparison. That attack relies on injecting a unique address cookie while the victim uses Tor and recognizing it later when the same node connects directly to the clearnet. Monero's address propagation breaks this assumption: nodes share whitelist addresses with both incoming and outgoing neighbors, so an injected cookie is likely to spread beyond the intended target and lose uniqueness.

An adapted address-cookie design is possible only under stronger assumptions, for example by assigning victim-specific IP:port pairs and waiting for the target to connect to them later. However, this remains scenario-dependent and can be confused by scanners or unrelated peers. \textit{ProxyMark} avoids this dependency by directly linking Monero Tor identifiers to IP addresses through watermarking, while using originated-transaction capture to bind those identifiers to transactions.

More broadly, these results show that Tor transport alone does not remove protocol-level leakage from Monero's P2P layer. \textit{ProxyMark} does not rely on breaking cryptographic transaction privacy; instead, it composes role-dependent peer-list behavior, asymmetric transaction forwarding, proxy selection rules, and traffic watermarking. The attack should therefore be interpreted as a practical feasibility result under the stated adversarial capabilities, not as a claim that every Monero-over-Tor transaction is automatically deanonymized. It also suggests that effective defenses should jointly harden Monero's P2P behavior and its use of Tor, rather than treating the anonymity network as a standalone shield.

\subsection{Mitigation}
We organize mitigations according to the three stages of \textit{ProxyMark}. Some defenses remove the underlying protocol signal used by the attack, while others mainly raise the adversary's cost or lower the attack success rate.

\smallskip
\noindent\textbf{For the node role identification method.}
Our method exploits that a Monero Tor hidden service node always appends its own onion address as the last entry in Timed Sync Response messages to its outgoing Monero Tor hidden service peers. A mitigation is to send the own onion address only once in the initial Handshake Response and exclude it from subsequent Timed Sync Response messages (with its \texttt{last\_seen} timestamp set to 0; see Appendix~\ref{appendix_onion_address_analysis_single}). This eliminates the repeated last-entry fingerprint that our method relies on, making the mitigation fully effective.
 
\smallskip
\noindent\textbf{For the originated transaction identification method.}
The root cause is that Monero Tor nodes forward originated transactions only to their outgoing Monero Tor hidden service peers, while relayed transactions are forwarded only to public peers. As a result, any transaction received by a Monero Tor hidden service node from an incoming peer over internal Tor is necessarily an originated transaction of that peer. The fundamental mitigation is to extend Dandelion++ to the anonymity network: with stem-phase forwarding over hidden service connections, a transaction received from an incoming internal Tor connection could be either originated or relayed in the stem phase, making them indistinguishable to the adversary. This requires protocol-level changes but eliminates the root cause entirely.
 
The outgoing connection occupation method and the proxy selection bias method are amplification techniques built on top of this root cause. For the former, Monero could verify onion address reachability before graylist insertion and limit the number of addresses accepted per Timed Sync message, raising the adversary's resource cost. For the latter, Monero nodes could cross-check reported block heights of outgoing Monero Tor hidden service peers against the public-peer-verified network height and exclude peers with significantly inflated heights from proxy selection. Both mitigations reduce the adversary's capture rate but are insufficient alone without addressing the root cause.
 
\smallskip
\noindent\textbf{For the watermark-based node location deanonymization method.}
Our method relies on the absence of a message frequency limit in Monero, allowing the adversary to encode watermark signals by modulating the number of request messages per time window. Enforcing a strict fixed-rate policy for Timed Sync Request, Ping, and Support Flags request messages (e.g., exactly one Timed Sync Request per minute) and disconnecting violating peers eliminates this encoding channel. Additionally, disallowing Timed Sync Request messages before handshake completion would neutralize the handshake-free watermark embedding method for Monero Tor hidden service nodes, as it forces the adversary to complete the handshake and reintroduces protocol-induced noise.

\section{Related Work}

We group related network-layer deanonymization work into attacks on ordinary cryptocurrency nodes and attacks on nodes communicating over Tor. Table~\ref{tab:rw_comparison} summarises the comparison with prior work.

\begin{table*}[t]
  \centering
  \caption{Comparison with prior network-level deanonymization solutions.}
  \label{tab:rw_comparison}
  \scriptsize
  \setlength{\tabcolsep}{4pt}
  \renewcommand{\arraystretch}{1.08}
  \begin{adjustbox}{max width=\textwidth}
  \begin{tabular}{>{\raggedright\arraybackslash}p{0.2\textwidth}
                  >{\raggedright\arraybackslash}p{0.2\textwidth}
                  >{\raggedright\arraybackslash}p{0.13\textwidth}
                  >{\centering\arraybackslash}p{0.05\textwidth}
                  >{\centering\arraybackslash}p{0.08\textwidth}
                  >{\raggedright\arraybackslash}p{0.17\textwidth}
                  >{\raggedright\arraybackslash}p{0.17\textwidth}}
    \toprule
    \multicolumn{1}{c}{\textbf{Work}} &
    \multicolumn{1}{c}{\textbf{Target}} &
    \multicolumn{1}{c}{\textbf{Main Vantage}} &
    \multicolumn{1}{c}{\textbf{Tor}} &
    \multicolumn{1}{c}{\textbf{Monero-Tor}} &
    \multicolumn{1}{c}{\textbf{Attribution Result}} &
    \multicolumn{1}{c}{\textbf{Gap / Scope}} \\
    \midrule
    Kaminsky~\cite{Kaminsky2011}; Koshy et al.~\cite{koshy2014} &
    Bitcoin-like public P2P &
    First relay / anomaly &
    \xmark &
    \xmark &
    Node-level heuristic &
    No Tor or Dandelion++ fit \\

    Biryukov et al.~\cite{biryukov2014,biryukov2019} &
    Bitcoin public P2P &
    Entry-peer fingerprint &
    \xmark &
    \xmark &
    Transaction--IP link &
    Needs stable entry peers \\

    Apostolaki et al.~\cite{apostolaki2021} &
    Public cryptocurrency P2P &
    AS/IXP traffic &
    \xmark &
    \xmark &
    Source inference &
    Requires unencrypted traffic \\

    Tramer et al.~\cite{tramer2020} &
    Zcash/Monero wallets &
    Wallet behavior &
    \xmark &
    Partial &
    Recipient IP &
    Wallet-side, fixed issues \\

    Eclipse attacks~\cite{heilman2015eclipse,shi2025eclipse} &
    Bitcoin/Monero public P2P &
    Connection hijacking &
    \xmark &
    Partial &
    Originated tx under eclipse &
    Requires stronger full-eclipse \\

    Biryukov~\cite{Biryukov2015} &
    Bitcoin users over Tor &
    Address cookie &
    \cmark &
    \xmark &
    IP under reuse &
    Needs clearnet reuse \\

    Gao et al.~\cite{gao2021} &
    Bitcoin Tor hidden services &
    Watermark + delay &
    \cmark &
    Partial &
    Hidden-service IP &
    Bitcoin forwarding model \\

    Tor hidden-services~\cite{biryukov2013trawling,tan2018toward,zhang2024hsdirsniper} &
    Tor hidden services &
    HSDir / eclipse &
    \cmark &
    Partial &
    Service exposure &
    No cryptocurrency tx binding \\

    Wang et al.~\cite{wang2024deanonymizing,wang2025time} &
    RPC users &
    Traffic--chain &
    \xmark &
    \xmark &
    User--IP link &
    RPC-specific setting \\

    Heimbach et al.~\cite{heimbach2025deanonymizing} &
    Ethereum validators &
    Attestation forwarding &
    \xmark &
    \xmark &
    Validator--IP link &
    Ethereum validator role \\
    \midrule
    \textbf{\textit{ProxyMark}} &
    \textbf{Monero over Tor} &
    \textbf{Proxy + watermark} &
    \textbf{\cmark} &
    \textbf{\cmark} &
    \textbf{Originated tx + source IP} &
    \textbf{Full Monero-Tor pipeline} \\
    \bottomrule
  \end{tabular}
  \end{adjustbox}
\end{table*}

\subsection{Ordinary-Node Attacks} 
Prior studies focused on transaction deanonymization and tracing~\cite{Kaminsky2011,koshy2014,biryukov2014,biryukov2019,apostolaki2021,tramer2020,kumar2023anonymity}, linking pseudonymous transfers to real-world entities and assessing privacy risks. 

Early attacks exploited transaction propagation patterns. Kaminsky~\cite{Kaminsky2011} observed that the first Bitcoin node forwarding a transaction is likely to be its originator, but this heuristic is weakened by Bitcoin's diffusion mechanism and Monero's Dandelion++ mechanism. Koshy et al.~\cite{koshy2014} leveraged anomalous relaying behavior, but such anomalies only appear in a small fraction of Bitcoin transactions. Biryukov et al.~\cite{biryukov2014,biryukov2019} used a node's outgoing peers as a fingerprint to associate its IP address with its originated transactions. However, this approach does not directly apply to Monero, because Monero nodes forward originated transactions to only one or two outgoing peers, rather than through a stable set of entry nodes.

Other studies consider stronger adversaries or different leakage channels. Apostolaki et al.~\cite{apostolaki2021} proposed an AS/IXP-level traffic analysis attack using unsupervised learning, but it requires observing unencrypted traffic and therefore does not apply to Tor-encrypted communication. Tramer et al.~\cite{tramer2020} exploited differences in how Zcash and Monero wallets handled transactions sent to themselves and transactions sent to other users. These differences allowed an attacker to test whether a wallet was the recipient of a transaction and thus reveal the recipient's IP address. However, the reported vulnerabilities have been fixed. Sharma et al.~\cite{kumar2023anonymity} proposed a Bayesian framework to quantify anonymity in P2P network anonymity schemes and showed that Dandelion and Dandelion++ may still provide limited anonymity under known topology. 

Eclipse attacks~\cite{heilman2015eclipse,wust2016ethereum,marcus2018low,tran2020stealthier,saad2023three,heo2023partitioning,li2023bijack,shi2025eclipse,shi2026eclipse} can also reveal a node's originated transactions by hijacking all of its connections, but they require a much stronger attack capability. 

Recent studies further show that network-layer leakage remains a practical concern in blockchain systems. Wang et al.~\cite{wang2024deanonymizing,wang2025time} correlated encrypted traffic with on-chain records to de-anonymize users relying on third-party RPC services; Heimbach et al.~\cite{heimbach2025deanonymizing} linked Ethereum validators to IP addresses through attestation forwarding behavior; and Kopyciok et al.~\cite{kopyciok2025friend} reported non-standard behavior among reachable Monero peers, suggesting active network monitoring.

\subsection{Tor-Node Attacks} 
De-anonymization over Tor is more challenging because Tor hides the network identifiers of cryptocurrency nodes. Biryukov et al.~\cite{Biryukov2015} de-anonymized Bitcoin users over Tor using address cookies, assuming that the same node sends sensitive transactions over Tor and later connects to Bitcoin without Tor on the same machine. This method can reveal the source IP address only under this specific usage scenario. Gao et al.~\cite{gao2021} studied Bitcoin Tor hidden service nodes and combined two steps: first, associating a hidden service's onion address with its IP address using signal watermarking and malicious Tor relays; second, identifying its originated transactions by delaying messages so that the attacker becomes the first receiver of those transactions. The second step targets Bitcoin's diffusion mechanism and does not directly apply to Monero, because Monero hidden service nodes only forward originated transactions to outgoing peers. Nevertheless, Gao et al.'s first step, namely watermark-based node location de-anonymization, is still relevant to Monero. In this paper, we build on this insight and address the Monero-specific challenges caused by its P2P protocol, Tor peer roles, and transaction forwarding rules.

\section{Conclusion}
We present \textit{ProxyMark}, a transaction deanonymization framework for Monero nodes operating over Tor. We show that Monero's current Tor integration can expose originated transactions because Tor nodes forward them through selected hidden service peers before clearnet propagation. We further design role specific watermarking methods to link Tor node identifiers with real IP addresses. We validate the attack pipeline on the live Tor network, Monero mainnet, and testnet, and we discuss protocol changes to transaction forwarding, peer selection, and message rate handling.

\bibliographystyle{IEEEtran}
\bibliography{bib}

\appendices

\section{Monero Node Initialization}
\label{appendix_initialization_process_of_Monero_nodes}

This appendix supplies the bootstrap details referenced in Section~\ref{subsec:p2p_messages}. Its purpose is to show how Section~\ref{sec-bck}'s public-peer and hidden-service peer lists are bootstrapped.

\begin{enumerate}
    \item[\textbf{a.}] \textbf{Ordinary seed lookup.} When a Monero node $A$ first joins the P2P network, it resolves hardcoded seed hostnames to obtain ordinary seed nodes. Node $A$ randomly selects one seed node $Seed_1$, sends a Handshake Request, receives a Handshake Response containing up to 250 ordinary node addresses, and then disconnects from $Seed_1$.
    \item[\textbf{b.}] \textbf{Public-peer connections.} Node $A$ randomly selects 12 ordinary peers $O_1,\ldots,O_{12}$ from the learned public peer list and establishes outgoing connections with them.
    \item[\textbf{c.}] \textbf{Hidden-service seed lookup.} If node $A$ is configured to use Monero hidden-service peers, it also connects to a hardcoded Monero Tor hidden service seed $Seed_2$, sends a Handshake Request, receives up to 250 hidden-service onion addresses, and disconnects from $Seed_2$.
    \item[\textbf{d.}] \textbf{Hidden-service peer connections.} Node $A$ then randomly selects 12 hidden-service peers $H_1,\ldots,H_{12}$ from the learned onion-address list and establishes outgoing hidden-service connections.
\end{enumerate}

Ordinary Monero nodes and Tor client nodes that connect only to public peers use steps a and b. Monero nodes over internal Tor that maintain hidden-service peers additionally use steps c and d.

\section{Tor Circuit Construction}
\label{appendix_tor_circuit_construction}

\begin{figure}[t]
    \centering
    \begin{subfigure}[t]{0.95\linewidth}
        \centering
        \includegraphics[width=\linewidth]{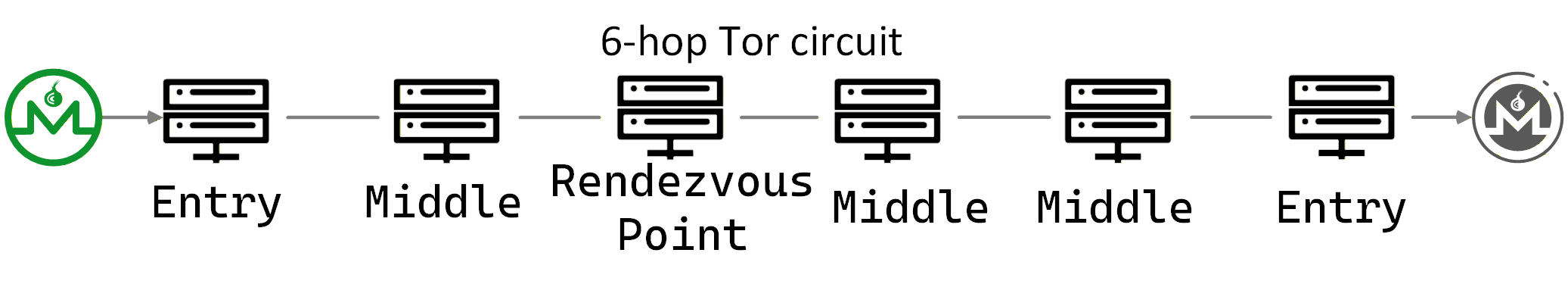}
        \caption{6-hop Tor path between a Monero Tor client node and a Monero Tor hidden service node.}
        \label{fig:appendix_6hop_tor_path}
    \end{subfigure}

    \vspace{0.8em}

    \begin{subfigure}[t]{0.75\linewidth}
        \centering
        \includegraphics[width=\linewidth]{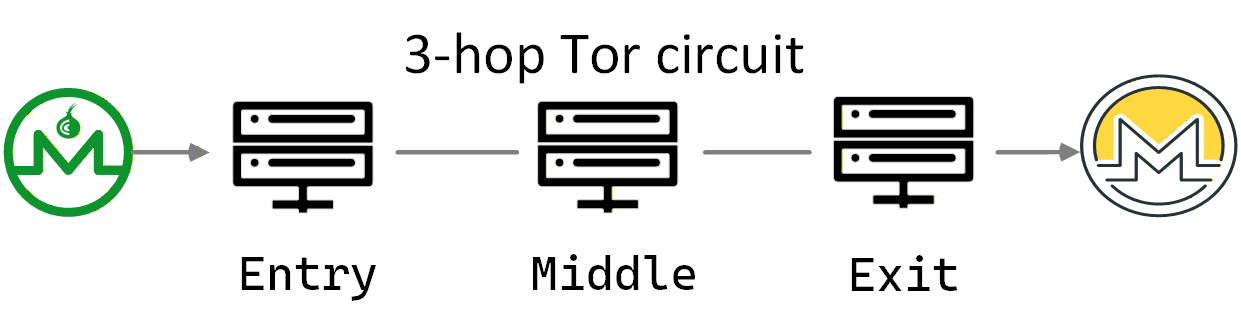}
        \caption{3-hop Tor circuit between a Monero Tor client node and an ordinary clearnet Monero node.}
        \label{fig:appendix_3hop_tor_circuit}
    \end{subfigure}

    \caption{Tor circuit structures for Monero node communication over Tor.}
    \label{fig:appendix_tor_circuit_construction}
\end{figure}

This appendix complements Figure~\ref{img_nodes_connection} by making the Tor-side paths explicit. Communication between a Monero Tor node and a Monero Tor hidden service node uses Tor's rendezvous design: the client side and the hidden-service side each build a 3-hop circuit to a rendezvous point, yielding a 6-hop Tor path. This applies to both Tor-client-to-hidden-service communication and hidden-service-to-hidden-service communication; Figure~\ref{fig:appendix_6hop_tor_path} uses the former as the example. By contrast, communication from a Monero Tor client to an ordinary clearnet Monero node uses a standard 3-hop Tor circuit with an entry relay, a middle relay, and an exit relay.

\section{Tor Cell Structure and Flow Control}
\label{appendix_tor_cell}

Section~\ref{subsec:tor_network_hs} only states that Tor uses fixed-size cells. The details relevant to our watermarking stage are the cell categories and the flow-control cells that may appear as encrypted traffic noise.

\begin{figure}[t]
  \centering
  \includegraphics[width=0.92\linewidth]{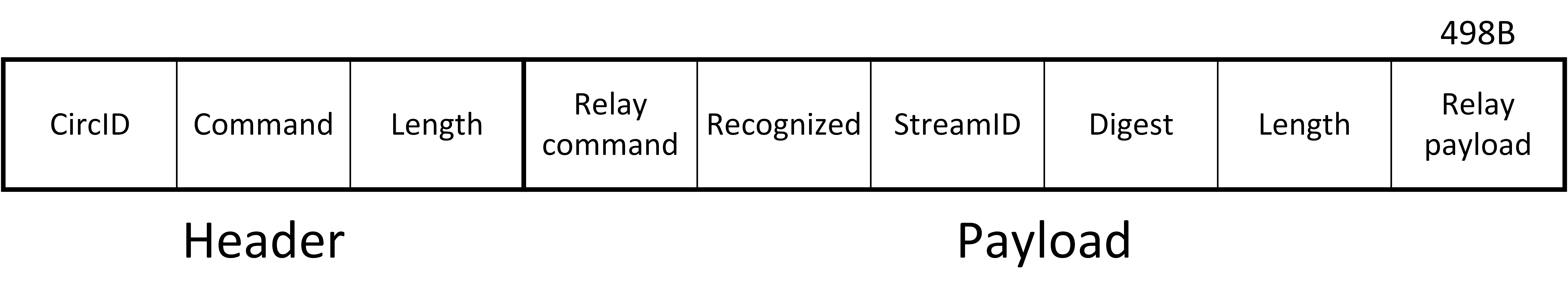}
  \caption{Composition of a Tor relay cell.}
  \label{img_tor_cell}
\end{figure}

Each Tor cell consists of a header, including fields such as \texttt{CircID}, \texttt{Command}, and \texttt{Length}, and an encrypted payload. The \texttt{CircID} identifies a circuit on a Tor link, while the \texttt{Command} field indicates the cell type, such as \texttt{CREATE}, \texttt{CREATED}, \texttt{RELAY}, \texttt{DESTROY}, or \texttt{RELAY\_EARLY}. Tor cells can be divided into control cells and relay cells. Control cells are used mainly to create and destroy circuits between adjacent hops. Relay cells carry end-to-end stream data through the circuit; a relay cell contains a relay header and a 498-byte relay payload inside the encrypted cell payload, as shown in Figure~\ref{img_tor_cell}.
Each relay on a circuit can inspect only link-level header fields and the timing, direction, and count of cells on its adjacent links; it cannot read the encrypted relay payload. Consequently, the watermark detector in Section~\ref{subsec-location-deanonymization} treats encrypted \texttt{RELAY} cells as indistinguishable counted units rather than attempting to parse their contents.

When a Tor user's application opens a TCP connection, the onion proxy opens a stream in an existing circuit. A \texttt{RELAY\_BEGIN} cell opens the stream, \texttt{RELAY\_CONNECTED} indicates success, \texttt{RELAY\_DATA} carries application data, and \texttt{RELAY\_SENDME} provides flow control. Each endpoint maintains package windows for circuits and streams, with initial values of 1000 and 500, respectively. Sending \texttt{RELAY\_DATA} decreases the relevant window, and receiving \texttt{RELAY\_SENDME} increases it by 50. A stream-level \texttt{RELAY\_SENDME} is sent after every 50 \texttt{RELAY\_DATA} cells on a stream, and a circuit-level \texttt{RELAY\_SENDME} is sent after every 100 \texttt{RELAY\_DATA} cells on a circuit. These flow-control cells are one source of background traffic considered in Section~\ref{subsec-location-deanonymization}.

\section{Single-Response Variant for Node Role Identification}
\label{appendix_onion_address_analysis_single}

Section~\ref{subsec-role-identi} uses three consecutive Timed Sync Responses as the conservative role-classification rule. A single-response variant is also possible because of how Monero handles the \texttt{last\_seen} timestamp.

Each address contained in a Handshake Response or Timed Sync Response message carries a \texttt{last\_seen} timestamp, which records when the responding node last communicated with that address. For topology protection~\cite{cao2020}, Monero normally advertises these timestamps as 0.

\smallskip
\noindent\textbf{Observation.}
Timed Sync Responses sent by a Monero Tor hidden service node to its outgoing hidden-service peers contain a distinctive exception. As illustrated in Figure~\ref{img_onion_analysis}, when a hidden service node $C$ responds to its outgoing hidden-service peer $D$, the last onion address is $C$'s own onion address and its \texttt{last\_seen} timestamp is set to the current time rather than 0.

\smallskip
\noindent\textbf{Variant.}
An adversarial Monero Tor hidden service node can therefore inspect a single Timed Sync Response received from an incoming peer. If the \texttt{last\_seen} timestamp of the last onion address is nonzero, the peer can be classified as a Monero Tor hidden service node, and the last address is inferred as its own onion address. The main pipeline uses three responses instead because repeated-address consistency is easier to validate experimentally and less sensitive to implementation-specific timestamp handling.

\section{Malicious Guard Selection Probability}
\label{appendix_malicious_guard_selection}

This appendix gives the probability derivation summarized in Section~\ref{sec-discuss}. Let $k$ be the number of malicious Tor relays injected by the adversary and $b$ be the bandwidth of each malicious relay. Following Tor's bandwidth weighting, let $B_g$, $B_e$, $B_{ge}$, and $B_n$ denote the total bandwidth of guard-only, exit-only, guard-exit, and neither-guard-nor-exit relays, respectively. The weighted bandwidth available for guard selection before malicious-relay injection is
\[
W_g = B_g + B_{ge} \cdot w_e,
\]
where
\[
w_e = \max\left(0, 1 -
\frac{B_g + B_e + B_{ge} + B_n}{3(B_e + B_{ge})}
\right).
\]
Using the Tor consensus on February 4, 2024, $w_e=0$ and $W_g=54{,}674.6875$ MB/s.

\subsection{Hidden Service Targets}
\label{section_probability_location_hidden}

Let $p_e$ be the probability that a target Monero Tor hidden service node selects at least one malicious Tor relay as its entry guard. Ignoring the small change in total available guard bandwidth after guard selection, we obtain
\[
p_e =
\frac{(k b)((k-1)b) + 2(k b)W_g}
     {(k b + W_g)^2}.
\]
This is the probability used for the hidden-service curve in Figure~\ref{img_tor_guard_probability_compare}.

\subsection{Tor Client Targets}
\label{section_probability_location_client}

Let $p_{ec}$ be the probability that a target Monero Tor client node uses a malicious relay as the entry relay of the circuit to a malicious hidden service node. A Tor client selects two guard relays and then uses one of them as the entry relay for a circuit. Under the same bandwidth approximation,
\[
p_{ec} =
\frac{(k b)((k-1)b) + 2(k b)W_g \cdot \frac{1}{2}}
     {(k b + W_g)^2}
\approx
\frac{k b}{k b + W_g}.
\]
This is the probability used for the client curve in Figure~\ref{img_tor_guard_probability_compare}.

\pgfplotsset{
  pmGuardAxis/.style={
    width=\linewidth,
    height=0.68\linewidth,
    xmin=0, xmax=1060,
    ymin=-0.015, ymax=0.50,
    axis lines=box,
    axis line style={draw=black, line width=0.5pt},
    tick align=outside,
    tick style={draw=black, line width=0.4pt},
    grid=major,
    major grid style={draw=gray!24, line width=0.25pt},
    xtick={0,200,400,600,800,1000},
    ytick={0,0.1,0.2,0.3,0.4,0.5},
    xlabel={Number of malicious relays ($k$)},
    ylabel={Selection probability},
    label style={font=\footnotesize},
    tick label style={font=\scriptsize},
    legend style={
      draw=none,
      fill=none,
      font=\scriptsize,
      at={(0.03,0.97)},
      anchor=north west,
      row sep=1pt,
    },
    clip=false,
  },
  pmServiceCurve/.style={
    pmGreen,
    line width=1.15pt,
    mark=square*,
    mark size=1.35pt,
    mark options={solid, fill=pmGreen},
  },
  pmClientCurve/.style={
    pmGold,
    line width=1.15pt,
    mark=triangle*,
    mark size=1.55pt,
    mark options={solid, fill=pmGold},
  },
}

\begin{figure}[t]
  \centering
  \begin{tikzpicture}
    \begin{axis}[pmGuardAxis]
      \addplot+[pmServiceCurve]
        coordinates {
          (10,0.01) (100,0.07) (200,0.13) (300,0.19)
          (400,0.24) (500,0.29) (600,0.33) (700,0.37)
          (800,0.40) (900,0.43) (1000,0.46)
        };
      \addlegendentry{Hidden service node ($p_e$)}

      \addplot+[pmClientCurve]
        coordinates {
          (10,0.00) (100,0.04) (200,0.07) (300,0.10)
          (400,0.13) (500,0.15) (600,0.18) (700,0.20)
          (800,0.23) (900,0.25) (1000,0.27)
        };
      \addlegendentry{Tor client node ($p_{ec}$)}

      \node[font=\scriptsize, text=pmGreen, anchor=west] at (axis cs:1000,0.46) {0.46};
      \node[font=\scriptsize, text=pmGold, anchor=west] at (axis cs:1000,0.27) {0.27};
    \end{axis}
  \end{tikzpicture}
  \caption{Probability that a Monero Tor node selects a malicious guard relay.}
  \label{img_tor_guard_probability_compare}
\end{figure}
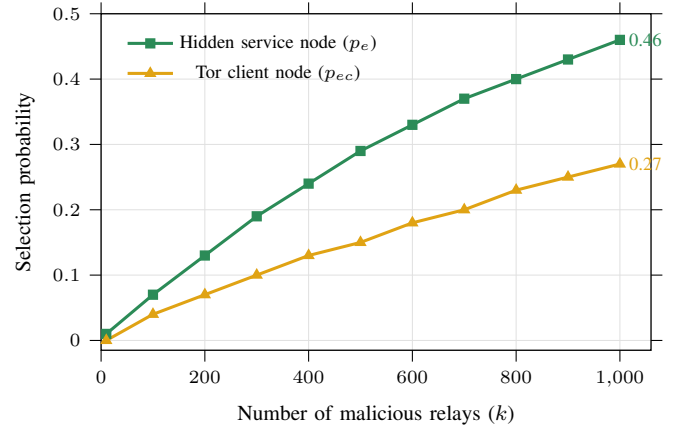

With $b=20$ MB/s and 179 malicious machines, the probabilities are $p_e=0.12$ for hidden service nodes and $p_{ec}=0.06$ for client nodes. Scaling to 1000 machines yields $p_e=0.46$ and $p_{ec}=0.27$, respectively.

\section{Feasibility of Biryukov's Approach in Monero}
\label{appendix_feasibilityof_Biryukov_in_Monero}

Section~\ref{sec-discuss} explains why \textit{ProxyMark} does not rely on Biryukov's Bitcoin-over-Tor address-cookie approach~\cite{Biryukov2015}. Here we give the protocol comparison in more detail.

\smallskip
\noindent\textbf{Anti-DoS prerequisite.}
Biryukov's approach depends partly on forcing selected peers into a constrained connection state. Monero has a comparable anti-DoS mechanism: each node maintains a penalty score for peers indexed by IP address, and a peer is banned for 24 hours once its score exceeds 10. In our test, 11 malformed Handshake Requests with an incorrect \texttt{network\_id} caused a Mainnet node $B$ to ban the sender address for 24 hours.

\smallskip
\noindent\textbf{Why the Bitcoin cookie does not transfer.}
The limiting factor is Monero's address propagation. Monero nodes maintain a graylist for learned addresses and a whitelist for addresses to which they have successfully connected. Handshake Responses and Timed Sync Responses share addresses selected from the whitelist, and these peer-list messages are sent to both incoming and outgoing neighbors.

Bitcoin's address-cookie attack exploits a rule that prevents forwarding an unsolicited ADDR message containing more than 10 addresses. The attacker can therefore inject an 11-address cookie into a Bitcoin victim while preventing the victim from forwarding that cookie. Monero does not provide the same forwarding barrier: once the injected addresses enter the victim's whitelist, they may be propagated to other neighbors. The original cookie can therefore lose uniqueness.

\smallskip
\noindent\textbf{Possible Monero adaptation and limitation.}
An adapted Monero cookie would need to satisfy two conditions: the injected addresses must correspond to online nodes that the victim can connect to, so that they enter the victim's whitelist, and the cookie must remain victim-specific even after the victim shares those addresses with neighbors. This would require victim-specific online addresses, such as a unique $address\_to\_N=\langle IP_1,port\_to\_N\rangle$ assigned only to victim $N$, with the corresponding node kept passive. A later connection to $address\_to\_N$ would likely identify $N$, but scanners, unrelated peers, or address propagation beyond the intended target can create ambiguity. Same-IP/different-port entries reduce the IP-resource requirement, yet the method remains scenario-dependent and less direct than \textit{ProxyMark}.

\end{document}